\documentclass[1--10]{ar2e}
\usepackage{ulem}
\begin{document}
\input epsf.tex
\input psfig.sty

\newcommand{\roughly}[1]%
        {\mathrel{\raise.4ex\hbox{$#1$\kern-.75em\lower1ex\hbox{$\sim$}}}}
\newcommand\lsim{\roughly{<}}
\newcommand\gsim{\roughly{>}}
\newcommand\CL{{\cal L}}
\newcommand\CO{{\cal O}}
\newcommand\half{\frac{1}{2}}
\newcommand\beq{\begin{eqnarray}}
\newcommand\eeq{\end{eqnarray}}
\newcommand\eqn[1]{\label{eq:#1}}
\newcommand\eq[1]{Eq. \ref{eq:#1}}

\jname{Annu. Rev. Nucl. Part. Sci.}
\jyear{2003}
\jvol{53}
\ARinfo{xxxxx}

\title{TESTS OF THE GRAVITATIONAL INVERSE-SQUARE LAW}

\markboth{ADELBERGER, HECKEL \& NELSON}{INVERSE-SQUARE LAW TESTS}

\author{E.G.\ Adelberger, B.R.\ Heckel, and A.E.\ Nelson
\affiliation{Department of Physics, University of Washington, Seattle, Washington
98195-1560; email: eric@gluon.npl.washington.edu; heckel@phys.washington.edu; anelson@phys.washington.edu}}

\begin{keywords}
gravitation, experimental tests of inverse-square law, quantum gravity, extra dimensions
\end{keywords}
PACS Codes 04.80.-y, 11.25.-w, 04.50.th

\date{\today}
\begin{abstract}
We review recent experimental tests of the gravitational inverse-square law and
the wide variety of theoretical considerations that suggest the law may break
down in experimentally accessible regions.
\end{abstract}
\maketitle

\section{INTRODUCTION}
\label{sec:intro}
\subsection{Background}
\label{subsec:background}
Gravitation was the first of the four known fundamental interactions to be
understood quantitatively and the first ``grand unification'' in  physics.
Isaac Newton's Theory of Universal Gravitation
connected terrestrial phenomena (the``falling apple") with astronomical
observations (the ``falling Moon" and Kepler's Laws). This theory stood
virtually unchallenged
until Albert Einstein developed his relativistic
theory of gravitation in 1917. Since then,
General Relativity has  passed all
experimental tests and is today the standard model of gravitation. 
Yet some three
centuries after Newton, gravitation remains one of the most puzzling
topics in physics. Recently, a completely unexpected and fundamentally new
gravitational property was discovered using distant Type Ia supernovae:
the apparent 
{\it acceleration} of
the Hubble expansion \cite{ri:98,pe:99}, which is as yet unexplained. Furthermore,
gravitation is not included, and in fact not 
{\it includable},
in the imposing
quantum field theory that constitutes the standard model of particle physics.

There is a broad consensus that the two standard models are incompatible. The strong, weak, and electromagnetic interactions are explained as results of
the quantum exchange of virtual bosons, whereas the gravitational interaction is
explained as a classical consequence of matter and energy curving spacetime.
Because
quantum field theories cannot describe gravitation and General Relativity
predicts an infinite spacetime
curvature at the center of a black hole, neither of these
two standard models is likely to be truly fundamental.

Connecting gravity with the rest of physics is clearly the central challenge of
fundamental physics, and for the first time we have
a candidate theory (string or M-theory) that may unify gravitation with
particle physics. But the remaining theoretical problems  have focused
attention on possible new phenomena that could show up as
deviations from the familiar inverse-square law (ISL)
of gravity, generally at length
scales less than a few millimeters, but sometimes also at astronomical or even
cosmological distances. We review these speculations in
Section \ref{sec:spec}.

Although it is conventionally assumed that the ISL should be valid for
separations from infinity to roughly the Planck length
($R_P=\sqrt{G\hbar/c^3} = 1.6\times 10^{-35}$ m), until a few years ago this assumption had
only been precisely tested for separations ranging from the scale of the
solar system down to a few millimeters. The reasons for this are obvious:
On the one hand, there are no independently known mass distributions on
length scales larger than the solar system, and on the other hand,
it is difficult to get enough matter in close enough proximity to
obtain a background-free gravitational signal at length scales smaller than 1 mm.
This contrasts strongly with Coulomb's Law (and its electroweak generalization),
which has been tested for separations down to
$10^{-18}$ m in $e^+e^-$
leptonic interactions at high-energy colliders \cite{coul}. 
Although Coulomb's Law
has not been experimentally
verified at  large length scales (relative to laboratory dimensions), a null-type
laboratory measurement looking for effects of the galactic electromagnetic
vector potential, $A$, rules out deviations due to a finite photon mass for length scales
up to $\sim 2 \times 10^{10}$ m \cite{la:98}.
\subsubsection{Parameterizations}
Historically, experimental tests of Coulomb's and Newton's 
inverse-square laws 
were used to set limits on violations that, for gravity, took the form
\begin{equation}
F(r) =G\frac{m_1~m_2}{r^{2+\epsilon}}~.
\end{equation}
{}From the perspective of Gauss's Law, the exponent 2 is a purely geometrical effect
of  three  space dimensions, so  this parameterization was not well-motivated
theoretically. Instead, it is now customary to interpret tests of the
ISL as setting limits on an additional Yukawa contribution to the familiar $1/r^2$
contribution, which in the gravitational case creates a potential
\begin{equation}
V(r) = -G \frac{m_1~m_2}{r}\left[1+\alpha \; e^{-r/\lambda}\right]~,
\label{eq:yukawa}
\end{equation}
where $\alpha$ is a dimensionless strength parameter and $\lambda$ is a
length scale or range. The Yukawa contribution is the static limit of
an interaction due to the exchange of virtual bosons of mass
$m_b=\hbar/(\lambda c)$, where $m_b$ is the boson mass; the Yukawa form is
also useful in other contexts (see Section~\ref{sec:extra}).

Some investigators (see, e.g., \cite{wi:98}) have considered the
possibility that a nonzero graviton
mass could lead to a ``pure Yukawa'' gravitational potential
$V(r)=-G m_1 m_2 e^{-r/\lambda}/r$, recognizing that this
phenomenological form does not have a
well-defined theoretical foundation (see ref. \cite{newref_giu02} 
for a different approach to the implications of a nonzero gravitational mass).
Others have considered power-law modifications to the ISL~\cite{fi:01}:
\begin{equation}
V(r) = -G \frac{m_1~m_2}{r}\left[1+\alpha_N \left(\frac{r_0}{r}\right)^{N-1}\right]~,
\label{eq:power law}
\end{equation}
where $\alpha_N$ is a dimensionless constant and $r_0$ corresponds to a new
length scale
associated with a non-Newtonian process. Terms with $N=2$ and $N=3$ may
be generated by the simultaneous exchange of two massless scalar and two massless
pseudoscalar particles, respectively~\cite{su:93,dr:53,mo:87a}, while  $N=5$ may be
generated by the simultaneous exchange of two massless axions~\cite{fe:98} or a massless
neutrino-antineutrino pair~\cite{fi:96}.

In this review, we focus on the
parameterization of Equation~\ref{eq:yukawa}; any experiment that detects a violation of the
ISL will indicate a strength, $\alpha$, and a length scale, $\lambda$, that
characterizes the violation. Once a violation is detected, it will become necessary to
determine the functional form of the violation.
The parameterization of Equation~\ref{eq:yukawa} has strong implications for
experimental tests of the ISL. Any one test of the law
necessarily covers a limited range of length scales. Suppose, for example,
one performs a Keplerian test, comparing the orbits of two
planets orbiting a common sun. Clearly, the test is insensitive
to values of $\lambda$ much less than the orbit radius of the inner planet.
It is also insensitive to ranges of $\lambda$ much larger than the orbit
radius of the outer planet because both planets simply feel a
renormalized Newton constant $\widetilde{G}=G(1+\alpha)$. Consequently,
a great variety of experiments is needed to effectively explore
a wide variety of length scales. This contrasts
with limits on Yukawa interactions from ``equivalence principle'' tests, where
a single experimental result for a composition-dependent acceleration difference
typically provides a constraint
on $\alpha$ for values of $\lambda$ ranging from the length scale of the attractor to
infinity (see, e.g., \cite{ad:91}).
\subsection{Scope of This Review}
This review concentrates on experimental tests of the ISL
at length scales of millimeters or less, and on the wide range of theoretical
developments
suggesting that new phenomena may occur in this regime. We also discuss
speculations about possible ISL violations at much larger
length scales that could have important cosmological implications.
A extensive review of experimental results
at longer length scales~\cite{fi:99} appeared in 1999; we
update it in
Section~\ref{sec:astro} below. A review of extra ``gravitational'' dimensions, with
emphasis on collider signatures, has recently appeared in this journal~\cite{hew:02}. A recent review of tests of the ISL from microns to centimeters is Reference \cite{newref_lon03}.
Our review is focused on work
done since 1995 and should be current as of January 2003.
An earlier review~\cite{ad:91} covered spin-dependent forces that we
do not consider here.
\section{THEORETICAL SPECULATIONS}
\label{sec:spec}
\subsection{Unifying Gravity with Particle Physics: Two Hierarchy Problems}

The two greatest triumphs of twentieth-century physics are general relativity and quantum mechanics. However, we do not currently know how to link these two
theories, or  how to do calculations  consistently in situations where both
gravity and quantum effects are important, such as
near the Big Bang and the cores of black holes.
Clearly  general relativity 
must be contained in  a more fundamental quantum
theory that would allow sensible  calculations even in extreme conditions.
However, attempts to quantize general relativity have been plagued with difficulties.
Although one can construct an effective quantum field theory of gravity and
particle physics that is sufficiently accurate for many applications,
the  theory is infamously ``nonrenormalizable" or nonpredictive---an infinite
number of free parameters are needed to describe  quantum effects  at arbitrarily
short distances to arbitrary precision.

All known nongravitational  physics is includable within  the standard model of
particle physics---a quantum field theory in which the weak and electromagnetic
interactions are unified into a single framework known as the electroweak theory.
Symmetry between the weak and electromagnetic interactions is manifest above a
scale  of roughly 100 GeV. This  unification scale, where the electroweak symmetry
is spontaneously broken, is known as the electroweak scale.  The electroweak scale
is set by a condensate of a scalar field known as the Higgs field that has a
negative mass-squared term of order $(100$~GeV$)^2$ in its potential.
All three forces of the standard model, the electromagnetic, weak, and strong
interactions, are similarly  unifiable into a simple group with a single coupling at the
fantastically high energy scale of $10^{16}$~GeV. This ``grand'' unified theory (GUT)
explains the quantization of electric charge and, provided there
exists a new symmetry between fermions and bosons known as supersymmetry, predicts the
observed value for the relative strengths of the weak and
electromagnetic couplings. But supersymmetry has not yet been observed in nature and,
if present, must be spontaneously broken.
Supersymmetry and GUTs are reviewed in
References~\cite{Ge:76,La:80,Sl:81,Ko:83,Ha:84,We:92,Dine:1996,Da:97,Ba:96,Lykken:1996,Ma:97,Polonsky:2001}.

Intriguingly, the Planck scale, $M_P=\sqrt{\hbar c/G}$,
at which quantum-gravity effects must become important,
$M_P c^2=1.2 \times 10^{19}$~GeV, is rather close to the apparent unification
scale of the other forces. This hints that all belong together in a unified framework
containing a fundamental scale of order $M_P$. Motivated by GUTs, the conventional
view is that the phenomenal weakness of gravity  at accessible energies---$10^{32}$
times weaker than the other forces at the electroweak scale---is due to the small
masses of observed particles relative to $M_P$.

In the standard model,  particle masses derive from the Higgs condensate.
The tremendous discrepancy between the scale of this condensate and the presumed
fundamental scale of physics  is  known as the 
{\it gauge-hierarchy problem}. 
In the minimal standard model, the smallness of the Higgs mass-squared parameter
relative to the
GUT  or Planck scales violates a principle
known as ``naturalness''---renormalized values of parameters that receive large
quantum
corrections  should not be much smaller than the size of the corrections. The Higgs
mass squared receives corrections proportional to the cutoff or maximum scale of
validity of the theory. Naturalness would therefore demand that to describe physics
at energies higher than about 1 TeV, the standard model should be contained within a
more fundamental theory in which the quantum corrections to the Higgs mass are
suppressed.
An example of   such a theory   is a supersymmetric extension of the standard model.
In  theories with spontaneously or softly broken supersymmetry, the  quantum
corrections
to scalar masses are proportional to the supersymmetry-breaking scale.
Provided the supersymmetry-breaking scale is of order 100 GeV, the electroweak scale
is natural, and the hierarchy question is why the supersymmetry-breaking
scale is so small compared with $M_P$. This latter problem is theoretically tractable;
in many supersymmetric models, the scale of supersymmetry  breaking is  proportional
to exponentially small, nonperturbative quantum effects \cite{Pop:98,Sh:99}.

A second, and much bigger, hierarchy problem is known as the 
{\it cosmological-constant problem}.
The strong observational evidence \cite{ri:98,pe:99} that the expansion
of the universe is accelerating can be explained by a nonvanishing cosmological
constant.
The concordance of cosmological data indicates \cite{Ca:00} that the universe is filled
with a
vacuum-energy density $\rho_{\rm vac}\sim 0.7 \rho_{\rm c}$, where $\rho_{\rm c}$ is
the critical density
$3 H^2 c^2/(8\pi G)$ and $H$ is the present value of the Hubble constant.
This gives $\rho_{\rm vac} \sim 4\ {\rm keV/cm}^3$, which corresponds to an
energy scale $\sqrt[4]{(\hbar c)^3 \rho_{\rm vac}} \approx$ 2 meV or a
length scale $\sqrt[4]{(\hbar c) /\rho_{\rm vac}}\sim 100\ \mu$m.  
Such a small energy density is particularly puzzling because the quantum corrections
to the vacuum energy density from particle physics scale as the  fourth power of the
cutoff of the effective theory.  Such a cutoff might be provided by new physics in
the gravitational sector. The  energy scale of new gravitational physics has been
presumed to be around $M_P$, which would imply a cosmological constant $10^{120}$
times larger than observed.  The success of the particle physics standard model at
collider energy scales is inconsistent with a cutoff lower than 1 TeV.
Even a
relatively low TeV cutoff gives a theoretical contribution to the cosmological constant
that is some $10^{60}$ times larger than experiment. References~\cite{Be:97} and~\cite{Su:97}
conjecture that this monstrous discrepancy could be eliminated with a much lower
cutoff for the gravitational sector of the effective theory, around 
1~meV, 
corresponding to new gravitational physics at a distance  of about 100 $\mu$m.
The theoretical framework for such a low  gravity scale is necessarily very speculative.
However, just as the gauge hierarchy  compels experimental exploration of the TeV scale,
the cosmological-constant problem strongly motivates submillimeter-scale tests of
gravity.

General relativity itself gives indications that the theory of quantum gravity is
radically
different from a conventional quantum field theory.
For instance, in
theories of gravity, the concept  of entropy must be generalized because entropy cannot
be an extensive quantity scaling like volume. In fact, strong evidence favors an upper bound on the entropy of
any region that scales as the surface area of the boundary of the region
\cite{Be:93,tH:93,Su:94}.
A further conjecture, the ``holographic principle," suggests that this entropy bound
indicates that the fundamental degrees of freedom of a gravitational theory can actually
be formulated in a lower-dimensional theory. Reference~\cite{Bo:02} reviews these ideas and their subsequent development.

M-theory is a popular candidate for a theory of quantum gravity.
This theory was called string theory when it was believed that its
fundamental degrees of freedom were one-dimensional objects propagating in a
10-dimensional spacetime. Six of these dimensions were assumed to be rolled up into a
compact manifold of size $\sim R_P$ and unobservable.
We now know that ``string''
theory  necessarily contains many types of objects, known as ``branes'' or ``$p$-branes,''
where $p$,
the number of spatial dimensions of the  $p$-brane, can be anywhere from $0$ to $9$.
This realization has revolutionized our understanding of
string theory. Furthermore,  string theory  is ``dual,''  or physically equivalent
as a quantum theory, to an 11-dimensional theory known as M-theory. There is much
theoretical
evidence that all known consistent string theories, as well as 11-dimensional
supergravity, are just weakly coupled limits in different vacua of a 
{\it single}
theory of quantum gravity.

{\it Extra} dimensions  might seem to contradict  the holographic assertion that
the fundamental theory is actually 
{\it lower}-dimensional.  However, as comprehensively
reviewed in Reference~\cite{Ah:99}, the
discovery that string theory on certain spacetimes with $n$ noncompact dimensions is
dual
to a nongravitational  gauge theory with $n-1$ dimensions provides additional
theoretical evidence  for  holography, as well as for string theory.
Strings, M-theory, p-branes, and duality have been reviewed extensively \cite{Sc:82,Dienes:1996du,Sc:96,Du:96,Po:96a,Po:96b,Ki:97,Va:97,Gr:98,Se:98,Ta:97,
Ba:98,Di:99,Jo:00,Dine:2000} and are the subject of several excellent
textbooks \cite{Gr:87,Pol:98}.

Until recently, it was believed that experimental verification of a theory of quantum
gravity
was out of the question, due to the impossibly short distance scale at which quantum
gravitational effects are known to be important. Furthermore, string theory contains a
stupendous number of vacua---with no known principle for selecting the one we should live
in---and so appears to have limited predictive power. Its chief phenomenological 
success to date is that in many of these vacua, the low-energy effective theory 
approximately resembles our
world, containing the fields of the standard model and  gravity propagating in four large
dimensions. A major unsolved difficulty is that all known  vacua are  supersymmetric,
although there are a variety of  conceivable ways  for the supersymmetry to be broken by
a small amount.

As we discuss below, although string theory makes no unique prediction,
all known ways of rendering our observations compatible with string theory
lead to new, dramatic signals in feasible experiments. In particular,
the discovery of branes has led to new
possibilities for explaining the gauge hierarchy and the cosmological constant. Many of
these can be tested in measurements of gravity at submillimeter scales, or in searches
for
small deviations from general relativity at longer distances.
\subsection{Extra Dimensions and TeV-scale Unification of Gravity}
\subsubsection{ ``Large'' extra dimensions}
\label{sec:extra}
It is usually assumed that the Planck scale is an actual physical scale, as is the weak
scale,
and that the gauge-hierarchy problem is to explain the origin of two vastly disparate
scales.
However, Arkani-Hamed, Dimopoulos  and  Dvali (ADD) \cite{Ar:98a} have proposed an
alternative explanation for the weakness of gravity
that has stimulated much theoretical and experimental work (see reviews in
\cite{hew:02,Ly:00,La:00,Ru:01,He:02a,Ue:02}).
Arkani-Hamed et al.\ conjecture that gravity is weak,
not because the fundamental scale is high but because gravity can propagate in new
dimensions
less than a millimeter in size. Such ``large'' new dimensions are not seen by the
standard-model
particles because these are confined to a 
three-dimensional subspace of the higher-dimensional theory.
Such a framework can be accommodated in string theory \cite{An:98}. A type of   $p$-brane
known as a
D$p$-brane does have gauge and other  degrees of freedom as light excitations that are
confined
to the brane. If the standard-model particles are all confined to such a D3-brane, we
will not
sense  additional dimensions except via their modification of the gravitational force
law.

The hierarchy problem can be reformulated in this framework.
One can assume that the fundamental scale $M_*$ is of order 1  TeV \cite{Ly:96}. There is
then no hierarchy between the weak scale and $M_*$ and
no gauge-hierarchy problem. If there are $n$ new dimensions, the higher-dimensional
Newton's
constant $G_{(4+n)}$ can be taken to be
\beq
G_{(4+n)}=\frac{4 \pi}{S_{(2+n)}}\left( \frac{\hbar}{M_* c}\right)^{(2+n)}\frac{c^3}
{\hbar},  \eeq
where $S_{(2+n)}$ is the  area of a unit $(2+n)$-sphere,
\beq
S_{(2+n)}={2 \pi^{(n+1)/2}\over \Gamma\left({n+1\over2}\right)}\ .
\eeq
At sufficiently short distances, the gravitational force at a separation $r$ would be
proportional
to $G_{(4+n)}/r^{2+n}$. To reconcile this with the  $1/r^2$ force law observed at long
distances,
Arkani-Hamed et al.\ take the $n$ new dimensions  to be compact.
At  distances that are long compared with the compactification scale,  the gravitational flux spreads
out
evenly over the new dimensions and is greatly diluted. Using Gauss's Law, one finds that
for
$n$ new dimensions with radius $R_*$,  compactified on a torus,  the effective Newton's
constant
at long distances is
\beq \label{add} G=\frac{\hbar c}{M_*^2}\left[\frac{\hbar}{M_* c}\right]^n \frac{1}{V_n}. \eeq
Here  $V_n$ is the volume  of the $n$-torus,  $(2\pi R_*)^n$. The relationship between
$R_*$ and $M_*$ for other geometries may be found simply by using the
appropriate formula for the volume.

The hierarchy problem is then transmuted into the problem of explaining the size of
the new dimensions, which are much larger than the fundamental scale. There are several
proposals for stable compactifications of  new dimensions that are naturally
exponentially large \cite{Ar:98b,Ar:99a,Co:99,Ch:01,Al:01}.

To test the ADD proposal directly, one should probe the ISL at a distance scale on the
order of
$R_*$. Compact new dimensions will appear as new Yukawa-type  forces, of range $R_*$,
produced by the exchange of massive spin-2 particles called Kaluza-Klein (KK)
gravitons \cite{Ka:21,Kl:26a,Kl:26b}.
To see this, note that the
components of the graviton momenta in the compact dimensions must be quantized. For
instance,
compactification of a flat  fifth dimension on a circle of radius $R$ would  impose the
condition
on $P_5$, the  fifth component of the graviton momentum,
$P_5=j\hbar /R$,
where $j$ is an integer.
The dispersion relation for a massless particle in five Lorentz-invariant dimensions is
\beq
E^2= \sum_{i=0}^3 c^2 P_i^2 + c^2 P_5^2\ .
\eeq
Comparing this with the four-dimensional massive dispersion relation
\beq
E^2= \sum_{i=0}^3 c^2 P_i^2 + c^4 M^2\ ,
\eeq
we see that the  fifth component of the momentum appears as a four-dimensional mass term.
A five-dimensional graviton thus appears as  an infinite number of new massive spin-2
particles.
For a flat new dimension compactified on a circle of radius $R$, the mass $m_j$ of the
$j$th
KK mode is
$m_j= j \hbar /(R c)$ with $j=1,2,...$.

In factorizable geometries (whose spacetimes are simply products of a
four-dimensional spacetime with an independent $n$-dimensional compact space),
the squared  wave functions  of
the KK modes are uniform in the new dimensions.
Low-energy effective-field theory analyses of the KK modes and their couplings
\cite{Gi:98,He:98,Ha:98,Mi:98} show that higher-dimensional general coordinate
invariance constrains this effective theory. Even at distances less than $R$,
KK mode exchange will not violate the equivalence principle. The
leading terms
in an expansion in $1/M_*$ contain a universal coupling of each graviton KK mode
$G^j_{\mu\nu}$ to the stress tensor of form
\beq
-\sqrt{{8\pi\over M_P}}\;\sum_j G^j_{\mu\nu}T^{\mu\nu},
\eeq
that is, each KK mode simply couples to the stress tensor in the same manner  as the
graviton.
To compute the correction to the ISL for nonrelativistic sources at long distances,
it suffices to consider the correction to the potential from the exchange
of the lightest KK gravitons.
The propagators for the KK states may be found in 
References~\cite{Gi:98,He:98,Ha:98,Mi:98}.

For $n$ new dimensions compactified on a flat torus, with the same radius $R_*$ for each
dimension, the lowest-lying KK mode  has multiplicity  $2n$  and Compton wavelength
$R_*$.
Direct searches for such new dimensions would  observe such KK gravitons  via the
contribution of their lowest-lying modes to the  Yukawa potential
of Equation~\ref{eq:yukawa}, giving $\alpha=8n/3$ and $\lambda=R_*$.
A factor of 4/3 occurs in $\alpha$ because a massive spin-2
particle has five polarization states, and the longitudinal mode does not
decouple from a nonrelativistic source
\footnote{Note that 
References~\cite{Ke:99} and \cite{Fl:99}
included a
contribution from a massless ``radion'' (gravitational scalar) in their Newtonian
potential, and the radion KK
modes in the Yukawa potential, leading to a  different value for
$\alpha$. We discuss the radion and why it should be
massive later in this section.}.
Other compact geometries will give
similar effects, although the value of $\alpha$ is quite model-dependent.

Assuming all new dimensions are compactified on a torus of  radius $R_*$, and $M_*=
1$~TeV,
Equation~\ref{add} gives
$$R_* \approx \frac{1}{\pi}10^{-17+{32\over n}}\, {\rm cm}\ .$$
The case $n=1$, $R_*= 3\times10^{12}$ m, is clearly ruled out. The case $n=2$, $R_*=0.3$
mm, is
inconsistent
with the results of Reference \cite{ho:01}. This case is even more
strongly
constrained by the observation of the neutrinos from supernova
1987A \cite{Ar:98c,Cu:99,Ba:99,Ha:00,Ha:01}. Gravitational radiation into the extra
dimensions
would rapidly cool the supernova before the neutrinos could be emitted, 
imposing a constraint
$R_*<0.7$~$\mu$m.  The extra gravitational degrees of freedom also necessarily spoil
the successful calculations of
big-bang nucleosynthesis unless $R_* <2$~$\mu$m,  
and the decay of the KK modes would add a
diffuse background of cosmological gamma rays whose non-observation implies
$R_*< 0.05~\mu$m \cite{Ha:99}.
For $n\ge 3$,  $R_*$ is less than about a nanometer, which is still allowed by
astrophysics,
cosmology, and direct searches.

It might, therefore, seem that direct observation of the new dimensions in ISL tests is
out of the question.
This conclusion is false. Astrophysical and cosmological  bounds are still
consistent
with a single extra dimension of size 1 mm---in such a scenario the hierarchy problem
might
be solved via the existence of several more much smaller new dimensions \cite{Ly:99b}.
Furthermore, as discussed in the next section, it is easy to alter Equation~\ref{add} 
and the
predictions for higher-dimensional graviton emission.
Finally, there is a strong argument
that the ADD proposal
should modify the ISL at a scale of order
$\hbar M_P/(c M_*^2)$.

In theories of gravity, the geometry of spacetime is dynamical and can fluctuate. In
particular,
the radius of new dimensions can fluctuate independently at each
point in our four-dimensional spacetime.
Thus, low-energy effective theories of compact extra dimensions inevitably contain
spin-0 fields parameterizing the radii of the new dimensions.
If the size of the
new dimensions is not determined by dynamics, then the linear combination of these
fields
that determines the extra dimensional volume is a massless Brans-Dicke
scalar \cite{newref_bra61} with gravitational strength coupling, known as the ``radion.''
A massless radion  is decisively ruled out by tests of general
relativity~\cite{Da:02}. Stabilization of the
volume of the extra dimensions is equivalent to
a massive radion.
Since, with a low fundamental scale, the  effective potential for the
radion should not be much larger than  $\CO(M_*^4)$, and its couplings are
proportional to $G_N$,  the radion  mass squared  should be lighter than 
$\CO(G_N M_*^4)$.
The radion will mediate a new, gravitational strength force,
with   $\alpha=n/(n+2)$ 
(\cite{An:02}; G.\ Giudice, R.\ Rattazzi, N.\ Kaloper,
private communications).
In many cases, the radion is  the lightest state associated with new dimensions.
For $M_*$  less than a few TeV, its range should be longer than  of order  100 $\mu$m. 
Even for relatively ``small'' new dimensions, with size of order an inverse TeV, 
the radion will, under certain assumptions, have a Compton wavelength in the  vicinity
of 100 $\mu$m~\cite{An:97,Ch:02}.
\subsubsection{Warped extra dimensions}
\label{sec:warped}
The previous discussion assumed the metric for the new dimensions is factorizable.
However, the most
general metric exhibiting four-dimensional Poincar\'e invariance is a ``warped product,''
\beq
ds^2=f(\xi_i)\eta_{\mu \nu} dx^{\mu} dx^{\nu} +g_{ij}(\xi_i) \xi_i\xi_j,
\eeq
where the  $\xi_i$ are the coordinates of the new dimensions, and $f$ and $g$ are
general
functions of those coordinates.  Solving the higher-dimensional Einstein equations for a
spacetime with an embedded
brane with nonvanishing tension typically requires warping. The ``warp factor''
$f(\xi_i)$ may be
thought of as a $\xi$-dependent gravitational redshift factor that
leads to a potential term in the graviton wave equation.
This potential can
have a dramatic effect on the $\xi$ dependence of the  wave functions of the
graviton, the graviton KK modes, and the radion.

Randall \& Sundrum~\cite{Ra:99a}  (RS-I)
noted that a large hierarchy can be obtained with a single small
new dimension if the metric takes the form
\begin{equation}
\label{rsi}
ds^2 = e^{- 2 k r_c \xi} \eta_{\mu \nu} dx^{\mu} dx^{\nu} + r_c^2 d \xi^2,
\end{equation}
where $\xi$ is a  coordinate living on the interval $[0,\pi]$, $k$ is a constant, and
$r_c$ is
the compactification scale.
This is just the metric for a slice of five-dimensional anti-deSitter
space (maximally symmetric spacetime with constant negative  curvature). It is also a
solution to the
five-dimensional Einstein equations with five-dimensional Newton's constant $1/M_*^3$ if there
is a
negative cosmological constant of size
$\Lambda = - 24 M_*^3 k^2$,
and if 3-branes are located at $\xi=0$ and $\xi=\pi$ with
tensions  $\pm 24 M_*^3 k$. A negative-tension brane seems unphysical,
but such bizarre objects can be constructed in string theory,
provided the spaces on each side of the
brane are identified with each other, that is, the brane represents a
boundary condition on the
edge of space. For large $kr_c$,  most of the extradimensional volume of this space is
near the
positive-tension brane at $\xi=0$.

To study the long-distance behavior of gravity in such a spacetime, one examines the
behavior of
small fluctuations of this metric of the form
\begin{equation}
\label{zeromodes}
ds^2 = e^{- 2 k r_c \xi} [\eta_{\mu \nu}+ {h}_{\mu \nu}(x)]
 dx^{\mu} dx^{\nu} +
r_c^2 d \xi^2 \ .
\end{equation}
Here $h_{\mu\nu}$ is the four-dimensional graviton.
Plugging this metric into Einstein's equations and linearizing in $h$, one finds $h$ is
a  zero mode,
or massless solution to the equations of motion, whose  wave function in the compact
dimension simply
follows the warp factor $e^{- 2 k r_c \xi}$.
Thus, there is a massless four-dimensional  graviton that
is localized about the brane at $\xi=0$ and exponentially weakly coupled to
matter on the brane at $\xi=\pi$. If we further hypothesize that the
latter brane is where the standard model lives, the
weakness of gravity is explained for a moderate value of $k r_c \sim 12$.
Both $k$ and $r_c^{-1}$ can be of the same order of magnitude as the fundamental scale,
and so there
is no large hierarchy in the input parameters.

As in the ADD case, the RS-I model  has a radion parameterizing the compactification
scale. Goldberger
\& Wise \cite{Go:99}  have shown that $k r_c$ in the desired range  can naturally be
stabilized
without large dimensionless inputs  if the theory contains a massive scalar that lives
in the bulk
and has source terms localized on the branes. The  radion then acquires a large mass of
order 100 GeV.
The curvature in the extra dimension has a huge  effect on the  KK graviton spectrum and
couplings.
The lightest KK modes have masses in the TeV region and large wave functions near our
brane, and therefore
$\CO(1)$ couplings to ordinary matter.
This model has unusual experimental signatures at colliders~\cite{hew:02} but is not
testable with
feasible probes of the ISL.

The RS-I model  teaches us that warping can have significant effects  on the
phenomenology of the new dimensions. The coupling strength and masses
of both the KK modes and the radion can be altered,
and the graviton can be localized, or bound to a brane.
Furthermore, warping is a generic phenomenon that should also occur in the ADD scenario.
Even a
very small amount of  warping can greatly alter the coupling of the
zero-mode graviton to our brane,
which makes this coupling either much stronger or much weaker than for the case of flat extra
dimensions \cite{Fo:00},  altering the relation of Equation~\ref{add}. Even in the case of
$M_*=1$~TeV and $n=2$,  with a very small amount of warping, the masses of the lightest
KK modes
can be either higher or lower than the inverse-millimeter scale predicted by the
unwarped case.
\subsection{Infinite-Volume Extra Dimensions}
\label{sec:infinite}
In a second paper~\cite{Ra:99b}, Randall \& Sundrum (RS-II) explored
the phenomenology  of a graviton  zero mode 
that is localized about a 3-brane embedded in a noncompact, infinite  extra dimension.
They found that five-dimensional gravity persists at all distance scales, with no
gap in the KK
spectrum, but at long distances the $1/r^2$ force, mediated by the
zero mode bound to the brane,  dominates, and
the extra dimension can be  unobservable at low energy. 
A simple model of this effect
is
given by the metric
\begin{equation}
\label{rsii}
ds^2 = e^{- 2 k |z|} \eta_{\mu \nu} dx^{\mu} dx^{\nu} + dz^2,
\end{equation}
where $z$, the coordinate of the  fifth dimension, is noncompact. This metric, which
represents
two slices of anti-deSitter space glued together at $z=0$,  also solves Einstein's
equations, given  a negative bulk cosmological constant $ - 24 M_*^3 k^2$, and a single 3-brane
at $z=0$ of
positive tension $24 M^3 k$.
The total gravitational potential  between two masses $m_1$ and $m_2$ separated by a
distance $r$
on the brane may be found by summing up the contributions of the bound-state mode and
the
continuum KK spectrum, which, for distance scales longer than $1/k$, gives
\begin{equation}
V(r) = G_N { m_1 m_2 \over r}\left(1+{1 \over r^2 k^2} \right)
\end{equation}
with $G_N = \hbar^2 k/M_*^3$.
The experimental upper bound on $1/k$ from $N=3$ terms in Equation~\ref{eq:power law}
has not been explicitly computed
but should be similar to the bound on the radius  of an extra dimension.
Therefore $M_*$ must be larger than about $10^9$~GeV, and there is still a gauge
hierarchy.
With two or more infinite new dimensions, and a graviton confined to our 3-brane, it is
possible
to lower $M_*$ to 1 TeV \cite{Ar:99b}. In such a scenario, the weakness of
gravity is due to the zero-mode graviton  wave function spreading over the extra
dimensions,
as in the ADD proposal, but the width of the wave function is set by the curvature scale
rather
than by the size of the dimension.
Empirically, the main distinction between such weak localization and a
large new dimension is that there is
no gap in the KK spectrum and the ISL is modified by additional power-law
corrections rather than by new Yukawa forces.

The RS-I  explanation of the weakness of gravity---we live on a brane, the
graviton is confined to a different, parallel brane and its wave function
here is small---can also be realized in infinite
extra dimensions \cite{Ar:99b,Ly:99a}. Lykken \& Randall studied
such a configuration with a single
extra dimension and  concluded that the weakness of gravity could be
explained without input of any
large dimensionless numbers.  The chief test of their scenario would be strong
emission of graviton KK modes at a  TeV collider. The continuum of KK modes
would modify the ISL, but their effect
would only be significant  for distances smaller than $\sim10$ fm.
\subsection{Exchange Forces from Conjectured New Bosons}
Even  if new dimensions are absent or small, the ISL can be modified at accessible
distance scales by the exchange of new spin-0 or spin-1 bosons; spin-0 bosons would
mediate an
attractive Yukawa force while spin-1 bosons give a repulsive modification. Here we
review some general
considerations that apply to new bosons, and motivations for considering their
existence.
\subsubsection{Scalars: general theoretical considerations}
In order for a scalar particle, $\phi$, to exert a coherent force on matter, it
must  have a Yukawa coupling to  electrons, to $u$, $d$, or $s$
quarks, to the square
of the gluon field strength, or to higher-dimension operators such as
certain four-quark operators.  The candidates of lowest dimension are
\beq
\frac{m_e}{f} \phi \bar e e\ ,\qquad \frac{m_d}{f}\phi\bar d d\
,\qquad \frac{m_u}{f}\phi\bar u u
\ ,\qquad \frac{1}{f}\phi G_{\mu\nu}^a G^{a,\mu\nu}\ .
\eqn{ops}
\eeq
When embedded in the standard model, these all arise from dimension-5
operators, hence the common factor of $1/f$, where $f$ has dimensions
of mass. We have assumed that all chiral-symmetry-breaking operators
should be proportional to  fermion masses. With this
assumption,  and with all of the above operators  present,
the gluon coupling will dominate the scalar coupling to
matter. Because the matrix element of $G^2$ in a nucleon is roughly the
nucleon mass, $M_N$, such an interaction would lead to a
Yukawa potential of the form
given in Equation~\ref{eq:yukawa}
with $\lambda=\hbar/(m_\phi c)$, where $m_\phi$ is the scalar mass
and $\alpha\simeq M_P^2/(4\pi f^2)$.

An interaction $(\phi/f)G^2$ produces radiative corrections to $m_\phi$.
In the standard model with cutoff $\Lambda$, one finds
\beq
\delta m_\phi \simeq \frac{\Lambda^2}{4\pi f} \lsim m_\phi \ .
\eqn{lam}
\eeq
The inequality expresses the requirement of naturalness.
For $f=M_P$ and  $m_\phi=2\times 10^{-4}$ eV, corresponding to a Compton
wavelength of 1~mm, naturalness implies
$\Lambda \lsim 5\ {\rm TeV}$. This scale $\Lambda$ approximately
coincides with the
scale at which naturalness of the  electroweak-breaking  sector
demands new physics. A scalar coupled more weakly would
correspond to a higher value for $\Lambda$.
\subsubsection{Forces from axion exchange}
\label{sec:axion}
A major loophole in the above arguments is that the
interactions between  matter and a new scalar may not arise from any of the
operators in Equation 15, but rather from nonperturbative QCD
effects.  This is the case for the pseudoscalar axion invented to explain
why strong interactions conserve $CP$ to high precision. A pseudoscalar particle
would normally not produce a Yukawa force between unpolarized bodies, but
instantons in the
presence of $CP$ violation induce a scalar Yukawa
coupling of the axion to matter
that melts away above $\Lambda_{\rm QCD}$. The softness of that coupling
makes the radiative correction to the axion mass insignificant.
However, a $CP$-violating scalar axion Yukawa coupling to matter scales roughly as $m_u
\bar\Theta_{\rm QCD}/f_a\simeq \bar\Theta_{\rm QCD}(m_u m_a)/(m_\pi f_\pi)$, where
$m_u\lsim 5$ MeV
is the $u$ quark mass, and $\bar\Theta_{\rm QCD}\lsim 10^{-9}$\cite{theta} is the strong $CP$-violating
angle.

Thus, for an axion mass $m_a=10^{-4}$ eV, the scalar axion coupling is
at most
about $ 10^{-4}$ times gravitational strength.
ISL tests with unpolarized bodies probe the square of this coupling, so they
are quite insensitive to the axion. On the other hand, monopole-dipole
tests \cite{mo:84}, which search for a $CP$-violating force between unpolarized and
polarized
bodies, are linear in the coupling and should be a more sensitive axion probe.
\subsubsection{Scalars: cosmological considerations}
A  light, weakly interacting particle cannot decay or annihilate within a Hubble time,
so its relic energy abundance must be equal to or less than that of the observed dark
matter.
However, the cosmology of scalars presents an important difficulty. A natural
potential for a scalar in an effective theory below a cutoff $\Lambda$ has
the form $V\sim \Lambda^4 \hat V(\phi/f)$, where $\Lambda\approx
\sqrt{m_\phi f}$, and $\hat V$ is an arbitrary function that
is assumed to contain no large dimensionless numbers.
If all scalar couplings are
proportional to $1/f$, then the scalar lifetime is of order
$ 4\pi f^2/m_\phi^3$,  essentially stable.  If at a temperature
$T\sim \Lambda$ the thermal average of the scalar potential energy is
$\langle V\rangle \sim T^4$, then the scalar field  would have a large
expectation
value, $\phi\sim f$. The infinite-wavelength component of this expectation
value will be frozen until the Hubble scale is of order $1/m_\phi$, and will
subsequently act like cold dark matter.
Assuming the standard-model spectrum and standard cosmology for $T<\Lambda$
(e.g., that the reheat temperature following inflation is above $\Lambda$), 
then an initial scalar energy density of $T^4$ at $T=\Lambda$ implies a ratio
today of the energy in cold scalars to the energy in baryons of order
\beq
\frac{\rho_w}{\rho_B} \simeq 2\times 10^8
\left(\frac{\Lambda}{M_N}\right)\ ,
\eeq
which is clearly unacceptable.

Cosmology with light scalars can be made acceptable by invoking  a very late stage of
inflation
with Hubble constant
$H$ less than or approximately $m_\phi$. 
Then $\phi$ rapidly evolves to the
minimum of its potential. Once inflation ends, the universe must
reheat
to a temperature $T_R$. However, the minimum of the scalar potential at $T_R$
does not coincide with the minimum today, due to the  tadpole generated by the
interactions of Equation
15  at finite temperature. One must therefore check that
coherent scalar oscillations are not regenerated during the reheating
process
after inflation. If  reheating  causes the minimum of the
potential to
change suddenly, relative to the oscillation
time (of order $10^{-13}$~s), then  regeneration of the scalar condensate can be
significant.
We are almost completely ignorant of both the late
inflationary mechanism and the timescale for reheating $t_R$, but a rough bound
on  $t_R$ may be estimated from the reheating temperature
using the sudden inflaton-decay approximation
\beq
t_r\sim (2/3)H^{-1}\sim (2/3)\left({M_P\over T_R^2}\right)
\left(\sqrt{90\over 8\pi^3 g_*}\right)\ .
\eeq
For $T_R \gsim 10$~MeV, which is necessary for standard big-bang
nucleosynthesis, $t_R\sim 3\times 10^{-3}$~s. Much  higher  reheat 
temperatures might be necessary to generate the baryon-number asymmetry. For example,
a reheat temperature of about 100 GeV corresponds to a reheat time of order
$t_R\sim 3\times 10^{-11}$~s.

Provided this timescale  is much longer than the
scalar oscillation time $\hbar/(m_\phi c^2)$,
the evolution of the
minimum of the  potential can take place adiabatically, injecting
little energy into the coherent mode.
The requirement of
such a late stage of inflation with acceptable reheating constrains
theories of particle physics near
the weak scale but does not rule out the existence of light scalars.
\subsubsection{Bosons from hidden supersymmetric sectors}
\label{sec:hidden}
As we discussed in Section \ref{sec:spec}, new physics is expected at the TeV scale.
One candidate for this new physics is supersymmetry, which  is expected in
unified theories, and which can explain  the gauge hierarchy. Unbroken supersymmetry
predicts an
unobserved degeneracy between fermions and bosons, so supersymmetry must be broken at
a scale
of 100 GeV or higher. The most popular scenario involves supersymmetry breaking at a
scale of $M_S\sim 10^{11}$ GeV in a ``hidden'' sector that couples to our visible
world only via
gravity and interactions of similar strength. The apparent scale of supersymmetry
breaking in the
visible world would then be of order $M_S^2/M_P\sim 10^3 $~GeV. In other scenarios,
supersymmetry breaking is communicated to the visible world by the gauge
forces of the standard model,
and the supersymmetry-breaking scale is as low as $M_S\sim10^4$~GeV.
The supersymmetry-breaking scale is linked to $m_{3/2}$,  the mass of the gravitino (the
spin-$\textstyle{\frac{3}{2}}$ superpartner of the graviton), through the relation $m_{3/2}=M_S^2/M_P$.
Well-motivated theoretical expectations for the gravitino mass range from  1 meV to
$10^4$~GeV. In some
scenarios \cite{Ve:99,Sc:00,Ba:02a,Ba:02b}, the gravitino mass may be
linked with the size of the cosmological constant inferred from the
supernova observations and should be about 1 meV.

If there are hidden sectors---particles coupled to the visible sector only via
gravitational
strength interactions---the apparent scale of supersymmetry breaking in those sectors
would
typically be of order $m_{3/2}$. Scalar particles from those
sectors could naturally have a mass in the meV range and  mediate gravitational strength
forces
with a range of about 100 $\mu$m.

Note that the severe cosmological problems typical of light weakly
coupled scalars discussed in the previous section  do not necessarily occur for a scalar
that is part of a hidden sector exhibiting
supersymmetry down to the meV scale. Such scalars might have a potential coming from
$\CO(1)$ couplings to  particles in the hidden sector, while maintaining a naturally small
mass and
gravitational-strength
couplings to particles in the visible sector. These couplings will allow for the scalar
field to relax to its minimum and for particle decay and annihilation.
\subsubsection{Forces  from exchange of stringy bosons}
\label{sec:stringy}
Supersymmetric hidden sectors are ubiquitous in  string theory.
All known acceptable vacua of  string theory are supersymmetric and contain a
tremendous number of
``moduli''---massless scalar fields whose expectation values set the parameters of the
effective
theory. These moduli are extremely weakly coupled, with couplings inversely proportional
to the
fundamental scale. In order to give these fields a mass, it is necessary to break
supersymmetry; however,
moduli necessarily couple weakly to the supersymmetry-breaking sector and, for a
low
supersymmetry-breaking scale, are expected to be extremely light.
Current understanding
is inadequate to predict the moduli masses, but a rough estimate suggests these should
be of order
$m_{3/2}$ \cite{DG:96}. The best way to look for moduli  is therefore to
test the ISL  at submillimeter distance scales.  The couplings of the moduli in any
given
vacuum are computable, and so there are definite predictions. The best-understood scalar
is the
dilaton, a modulus that determines the strength of the gauge couplings.
Its couplings to ordinary
matter can be determined and are nearly free of QCD uncertainties, so its discovery
could provide a
genuine smoking gun for string theory \cite{Ka:00a}.
\subsubsection{Forces from the exchange of weakly coupled vector bosons}
\label{spinone}
A new 
repulsive Yukawa interaction would be a signal for the exchange of a massive
spin-1 boson,
presumably a gauge particle.
In the ADD scenario, any gauge fields that propagate in the bulk of the new dimensions
would have
their couplings diluted by the same volume factor as the graviton and so would mediate a
force with
similar strength. Actually, since the gravitational force is also weakened by the
smallness of the
$M_N$ relative to $M_*$, one would expect any such gauge forces to be  stronger than
gravity
by a very large factor of $(M_*/M_N)^2\sim 10^6$--$10^8$. This is acceptable if the
range is
substantially shorter than  1 mm (see Section~\ref{sec:casimir results}).
Gauge bosons could have a mass in an interesting
range if the symmetry is broken via a scalar condensate on a brane.  The resulting
mass would be diluted by the bulk volume as well, and would naturally be in the range
$M_*/(VM_*^3)\sim M_*^2/M_P$. For $M_*$ of order a few TeV, the range would be about  100 $\mu$m \cite{Ar:98c}.
If the symmetry breaking occurs on the brane we live on, the  gauge-boson couplings to
standard-model matter could be substantially suppressed \cite{Dv:00c}.

Compactifications of string theory and other extradimensional theories often  contain
new massless
spin-1 particles,  known as graviphotons, that arise from components of the higher-dimensional graviton.
These  generally  do not couple to ordinary light matter, but
such bosons might acquire small masses and small,
gravitational-strength couplings to ordinary matter, e.g., by mixing with other  vector
bosons \cite{Bars:86,It:86,Barb:86,Barr:86}.
Light spin-1 bosons do not suffer from the naturalness or cosmological difficulties of
scalar
particles, provided that they  couple to conserved currents. However, spin-1 (and spin-0) boson
exchange necessarily ``violates'' the equivalence principle and the couplings of bosons
with masses less than
$1~\mu$eV are strongly constrained  by the experiment of Reference \cite{sm:00}.
\subsection{Attempts to Solve the Cosmological-Constant Problem}
\label{sec:lambda}
Comprehensive reviews of the cosmological-constant problem and the many attempts to
solve it are available \cite{We:88,Ca:91,We:00,Ca:00,Wi:00}. Recent theoretical activity on
this topic has been intense but is still inconclusive.
Here we simply mention a
few of
the interesting recent proposals that imply modifications of the ISL at {\it long} distances.

Beane \cite{Be:97} pointed out that in any local effective quantum field theory,
naturalness would
imply new gravitational physics at a distance scale of order 1~mm that would cut
off
shorter-distance contributions to the vacuum energy. Sundrum \cite{Su:97} has
speculated about the
sort of effective theory that might do this. Sundrum proposed that the graviton is  an
extended
object, with size of order  1~mm, and has been exploring how to construct a
natural  and
viable effective field theory from this picture \cite{Su:97,Su:03}. It is still
not clear
how self-consistent this effective theory is, but it does have the great virtue of
making a
definite, testable experimental prediction---gravity should shut off below a distance
scale of
order 100 $\mu$m.

Many people have attempted to use extra dimensions to explain the smallness  of the
cosmological constant,
motivated by the alluring observation \cite{Ru:83} that in higher-gravitational theories
with
branes,  the four-dimensional vacuum energy or brane tension  does not necessarily act as a
source
of four-dimensional gravity but can instead lead to curvature only in the new dimensions.
So far
no solved, consistent example actually yields a small  cosmological constant in the
four-dimensional effective description without extreme fine tuning or other problematic
features.

Theories with branes  and noncompact new dimensions allow another surprising
phenomenon
known as quasilocalization of gravity \cite{Gr:00,Kar:00,Dv:00a,Dv:00b}. In these
theories, as in
RS-II, long-distance gravity is higher-dimensional. However, there is no zero mode
bound to our 3-brane. There is, instead, a metastable quasibound state that
propagates four-dimensionally
along the brane over times and distances that are short compared with some maximum scale.  The ISL,
and four-dimensional general relativity, will approximately apply
from $r_{\rm min}$ to $r_{\rm max}$, but not to arbitrarily long distances.
The consistency of various  theories of quasilocalization
is still under debate and the theories themselves have been mutating rapidly.

The holographic principle insinuates
that a local description of a gravitational theory must break down somehow, because there
are not
enough degrees of freedom to allow independent observables at different spacetime
points.
Several theorists have speculated that the breakdown of locality might even occur in a
subtle way at astronomical or even longer distances, and that this might explain the
size of the
cosmological constant \cite{Ba:95,Ho:97,Co:98,Ba:00,Th:02}. In the Banks scenario \cite{Ba:95,Ba:00,Ba:02a,Ba:02b},
supersymmetry ends up being broken at a scale of a few TeV by nonlocal effects due to
the cosmological constant, leading to masses for the gravitino,
dilaton, and other moduli of order 1 meV and deviations from the ISL at 100 $\mu$m.

Many of the above ideas share the possibility that there is some scale  $r_{\rm max}$
beyond which
Einstein  gravity is modified. Modifying gravity at long distance allows  a new
approach to the cosmological
constant. The observed acceleration of the universe might be caused by a change in
the behavior of gravity at  the Hubble scale, instead of by  dark energy \cite{De:01}.
The fascinating prospect that  the effective Newton's constant might be strongly scale-dependent at
large
distance scales (gravity as a ``high-pass spatial filter'') leads to
a
new view of the cosmological-constant
problem. Conventionally, it is assumed that the vacuum energy gravitates so weakly
because,
for some mysterious reason, this energy is actually very small. But if the strength of
gravity
depends on the wavelength of the source, it becomes credible that the vacuum energy is
indeed very large but that it gravitates weakly because
it is very smooth.
Ideas along these lines have been pursued  \cite{Dv:02a,Dv:02b,Ar:02}.

References~\cite{Dvali:2001gm,Deffayet:2001uk,Gruzinov:2001hp,Lu:02,Dv:02c} 
present an
intriguing assertion about
theories of quasilocalization that may account for the acceleration of
the universe. For any localized gravitational source,
there exists a distance scale $r_*$ beyond which the graviton will acquire
an extra polarization state that couples to the source so that
the strength of gravity changes. This scale $r_*$ is a function of the gravitational radius of the source and $r_{\rm max}$; it decreases for less massive
gravitating objects. Dvali et al.~\cite{Dv:02c}
argue that ultraprecise measurements of the anomalous precession of the
perihelion of planetary orbits can test models of quasilocalization that
explain the cosmological
acceleration. For instance,  a 17-fold  improvement in this measurement
in the Earth-Moon system via lunar-laser ranging (LLR) would test a particular model in
Reference~\cite{Dv:02c}.
\section{EXPERIMENTAL CHALLENGES}
\label{sec:exp}
\subsection{Signals}
The dominant problem in testing gravitation at short length scales is the extreme
weakness of gravity. This forces the experimenter to adopt
designs that maximize the signal and minimize backgrounds and noise.
For example, one
could measure the force between spheres  \cite{mi:88}, between
cylinders
\cite{ho:85,ed:00},
between a sphere and a plane 
\cite{la:97,umo:98}, 
or in planar geometry 
\cite{ho:01,lo:02}.
Clearly, at a given minimum separation, the signal from a short-range
interaction, per
unit test-body mass, is least for two spheres and greatest for two planes.

The Yukawa force between two spheres of radii $r_1$ and $r_2$ and
masses $m_1$ and $m_2$, whose centers are separated by $s$, is
\begin{equation}
F_Y=\alpha G~m_1 m_2  \Phi \left( \frac{r_1}
{\lambda}\right)\Phi  \left( \frac{r_2}{\lambda}\right)
\left( 1 + \frac{s}{\lambda}\right) \frac {e^{-s/\lambda}}{s^2}~,
\end{equation}
where $\Phi(x) = 3(x \cosh x - \sinh x)/x^3$. For $x \gg 1$,
$\Phi(x) \approx 3 e^x/(2x^2)$, whereas for $x \ll 1$,
$\Phi(x) \approx 1$.
Therefore, for $\lambda \ll r$, the ratio of Yukawa to Newtonian forces
for two spheres of radius $r$ separated by a gap $d$ is
\begin{equation}
\frac{F_Y}{F_N} \approx \alpha \frac{9}{2} \frac{\lambda^3}{r^3} \left(1+\frac{d}{2r}\right)
e^{-d/\lambda}~.
\end{equation}

The potential energy from a Yukawa interaction between a flat plate
of area $A_p$, thickness $t_p$, and density $\rho_p$ at distance $d$
from an infinite plane of
thickness $t$ and density $\rho$ is
\begin{equation}
V_Y=2 \pi \alpha G \rho_p \rho \lambda^3 A_p \left[1-e^{-t_p/\lambda}\right]
\left[1-e^{-t/\lambda}\right] e^{-d/\lambda}~,
\label{eq:plate yukawa potential}
\end{equation}
if end effects are neglected. The corresponding force is
\begin{equation}
F_Y=2 \pi \alpha G \rho_p \rho \lambda^2 A_p \left[1-e^{-t_p/\lambda}\right]
\left[1-e^{-t/\lambda}\right] e^{-d/\lambda}~.
\label{eq:yukawa force of planes}
\end{equation}
In this case, for $\lambda$ much less than the thicknesses,
the force ratio becomes
\begin{equation}
\frac{F_Y}{F_N} \approx \alpha \, \frac{\lambda^2}{t_p t} \, e^{-d/\lambda}~.
\end{equation}

The potential energy of a Yukawa interaction between a sphere of radius $r$
and mass $m$ above an infinite plane of thickness $t$ and density
$\rho_p$ is
\begin{equation}
V_Y=\pi \alpha G m \rho \lambda^2  \Phi(r/\lambda) e^{-s/\lambda},
\end{equation}
where $s$ is the distance from the center of the sphere to the plane.
The corresponding force is
$F_Y=\pi \alpha G m \rho \lambda \Phi(r/\lambda)e^{-s/\lambda}~.$
In this case, for $\lambda \ll r$, the force ratio becomes
\begin{equation}
\frac{F_Y}{F_N} \approx \alpha \, \frac{3}{4} \, \frac{\lambda^3}{r^2 t} e^{-d/\lambda},
\end{equation}
where $d$ is the gap between the spherical surface and the plane.
\subsection{Noise Considerations}
\label{sec:low-f noise}
Thermal noise in any oscillator sets a fundamental limit on the
achievable statistical error of its amplitude.
A single-mode torsion oscillator subject to both velocity and internal
damping obeys the equation
\begin{equation}
{\cal T}=I\ddot{\theta} + b \dot{\theta} + \kappa (1+i\phi) \theta~,
\end{equation}
where ${\cal T}$ is the applied torque, $I$ the rotational inertia, $\theta$
the angular deflection of the oscillator, and $\kappa$ the torsional spring
constant of the suspension fiber. The velocity-damping coefficient $b$ accounts
for any losses due to viscous drag, eddy currents, etc., and
the loss angle $\phi$ accounts for internal friction of the suspension fiber.
We compute the spectral density of thermal noise following
Saulson's \cite{sa:90} treatment based on the fluctuation-dissipation
theorem. The spectral density of torque noise power (per Hz)
at frequency $\omega$ is
\begin{equation}
\langle {\cal T}^2_{{\rm th}}(\omega)\rangle=4 k_B T \Re (Z(\omega)),
\end{equation}
where $k_B$ is Boltzmann's constant, $T$ the absolute temperature, and
$Z={\cal T}/\dot{\theta}$ the mechanical impedance.

First consider the familiar
case of pure velocity damping ($b > 0$, $\phi = 0$) where
$Z(\omega)=iI\omega+b+\kappa/(i\omega)$. In this case, the spectral
density of torque noise,
\begin{equation}
\langle{\cal T}^2_{{\rm th}}(\omega)\rangle= 4k_B T \, \frac{I \omega_0}{Q}
\end{equation}
($\omega_0=\sqrt{\kappa/I}$ is the free resonance frequency and
$Q=I \omega_0 /b$
the quality factor
of the oscillator), is
independent of frequency. The corresponding spectral density of angular-deflection noise
in
$\theta$ is
\begin{equation}
\langle\theta^2_{{\rm th}}(\omega)\rangle=\frac{4 k_B T}{QI}\frac{\omega_0}
{(\omega_0^2-\omega^2)^2+
(\omega_0\omega/Q)^2}~.
\label{eq:theta noise for velocity damping}
\end{equation}
Note that the integral of Equation~\ref{eq:theta noise for velocity damping} over all $f=
\omega/(2 \pi)$
is $k_B T/\kappa$, consistent with the equipartition theorem.
The signal due to an external torque ${\cal T}$ is
\begin{equation}
|\theta(\omega)|=\frac{{\cal T}}{I} \frac{1}{\sqrt{(\omega_0^2-\omega^2)^2+
(\omega \omega_0/Q)^2}},
\end{equation}
so  the signal-to-noise ratio in unit bandwidth has the form
\begin{equation}
S(\omega)=\frac{|\theta(\omega)|}{\sqrt{\langle\theta^2_{{\rm th}} + \theta^2_{\rm ro}}
\rangle}=
\frac{{\cal T}}{\sqrt{4 k_B T \omega_0 I/Q + \langle \theta^2_{{\rm ro}}\rangle
I^2((\omega_0^2-\omega^2)^2+(\omega \omega_0/Q)^2)}},
\end{equation}
where we have included a noise contribution $\langle \theta^2_{{\rm ro}}\rangle$ from
the
angular-deflection readout system.
The signal is usually placed at a
frequency $\omega \le \omega_0$ to avoid attenuating the deflection
amplitude $\theta$ because of oscillator inertia.

Now consider the case of pure internal damping ($b=0$, $\phi >
0$) where
$Z=iI\omega + \kappa/(i\omega) +\kappa \phi/\omega$. In this case, the
spectral density of thermal noise has a $1/f$ character,
\begin{equation}
\langle {\cal T}^2_{{\rm th}}(\omega) \rangle= 4k_B T \, \frac{I \omega_0^2}{\omega Q},
\label{eq:torque noise}
\end{equation}
where now $Q=1/\phi$. The corresponding spectral density of thermal noise in the angular
deflection is
\begin{equation}
\label{eq:thermal}
\langle \theta^2_{{\rm th}}\rangle=\frac{4k_b T}{Q \omega I}
\frac{\omega_0^2}{(\omega_0^2-\omega^2)^2+(\omega_0^2/Q)^2}~.
\end{equation}
The signal-to-noise ratio in unit bandwidth is
\begin{equation}
S=\frac{{\cal T}}{\sqrt{4k_B T I \omega_0^2/(Q \omega)+
\langle \theta_{\rm ro}^2\rangle I^2((\omega_0^2-\omega^2)^2+(\omega_0^2/Q)^2)}},
\end{equation}
so  it is advantageous to boost the signal frequency above $\omega_0$
until $\theta^2_{\rm ro}$ makes a significant contribution to the noise.
\subsection{Backgrounds}
Electromagnetic interactions between the test bodies are the primary source of
background signals and may easily dominate the feeble gravitational signal.
In the following sections, we discuss
the dominant electromagnetic background effects in ISL experiments.
\subsubsection{Electric potential differences and patch fields}
\label{sec:patch}
Electric charges residing on insulating or ungrounded test bodies are difficult to
quantify, and 
Coulomb forces acting on such bodies can exceed their weights. 
For this reason, ISL tests typically employ conducting grounded test bodies.
Even so, a variety of effects can give the test bodies different electric potentials.
If dissimilar materials are used for the test bodies, a potential difference equal to
the difference between the work functions of the
two materials is present, typically of order  $1$ V. Even if the same material is used
for both test
bodies or the test bodies are both coated with the same material, such as gold, small
differences in the
contact potentials connecting the test bodies to ground can leave a net potential
difference between
the test bodies. With care, such contact potential differences can be reduced to the
level of a few mV \cite{ha:00}).

Neglecting edge effects, the attractive
electric force between a conducting plate with area $A$ parallel to an infinite
conducting plate is
$F_E(d) = \epsilon_0 A V^2/(2d^2)$,
where $d$ is the separation between the plates, $V$ is the potential difference between
the plates, and
$\epsilon_0$ is the permitivity of free space. For 1-mm-thick plates with a density of
$10$ g/cm$^3$,
separated by $0.1$ mm, $F_E$ becomes as large as $F_N$ for a potential difference of 10
mV,
and the electric
force grows with decreasing separation whereas the Newtonian force is constant.

Even if test bodies are at the same average potential, they experience a residual
electric
interaction from patch fields---spatially varying microscopic electric potentials
found on the surface of materials \cite{sp:96}. Patch fields arise because different
crystal planes of a
given material
have, in general, work functions \cite{mi:77} that can in extreme cases differ by as much as 1 V.
To the extent that the surface is a mosaic of random
microscopic crystal planes, local potential differences will occur with a scale size
comparable to the
size of the microcrystals. For example, the work functions of different planes of W crystals differ by 0.75 V. Gold is a good choice for test-body coating because the work functions of its
crystal
planes vary by only 0.16 V. Surface contaminants also contribute to the local variation
of the electric
potential, altering the local work function and providing sites for the trapping of
electrical charge. In the limit that the patches are smaller than the separation, the
patch field
force \cite{sp:96} scales as $1/d^2$.
\subsubsection{Casimir Force}
\label{sec:casimir}
Vacuum fluctuations of the electromagnetic field produce a fundamental
background to ISL tests at short length scales.
The Casimir force \cite{hbc:48} between
objects in close proximity may be viewed as arising either from the
modification of the boundary conditions for zero-point electromagnetic
modes or from the force between fluctuating atomic dipoles induced by
the zero-point fields \cite{eml:56}. The Casimir force
can be quite large compared to the force
of gravity. The Casimir force between two grounded, perfectly conducting,
smooth, infinite planes at zero temperature, separated by a distance $d$, is
attractive with a magnitude of
\begin{equation}
\frac{F_C}{A}=\frac{\pi^2 \hbar c}{240 d^4}~.
\label{eq:cas1}
\end{equation}
For a $1$-mm-thick plate of area $A$ near an infinite plate of thickness $1$ mm (again,
both with
density $10$ g/cm$^3$), $F_C$ becomes equal to $F_N$ at a separation of $d=13$ $\mu$m.

Because precisely aligning two parallel planes is so difficult, experimenters usually measure the force between a sphere (or
spherical lens)
and a plane. Assuming perfectly conducting, smooth bodies at zero temperature, the
Casimir force is attractive with a  magnitude of
\begin{equation}
F_C=\frac{\pi^3 R \hbar c}{360 d^3},
\label{eq:cas2}
\end{equation}
where $R$ is the radius of the sphere and
$d$ is the minimum separation between the surfaces of the sphere and plane. For a 1-mm-radius sphere
near an infinite 1-mm=thick plane (both with a density of 10 g/cm$^3$), $F_C$ becomes
equal to
$F_N$ at a  separation of  $d=2.5$ $\mu$m.

The Casimir-force expressions in Equations~\ref{eq:cas1} and \ref{eq:cas2} must be corrected
for
finite temperature, finite conductivity, and surface roughness (see below). All these
corrections vary with the separation, $d$, making it difficult to distinguish a
gravitational anomaly from an electrical effect.

\subsubsection{Electrostatic shielding}
Fortunately, backgrounds from  the Casimir force, electric potential differences, and patch-effect
forces can be greatly reduced
by using a moving attractor to modulate the signal on a stationary detector
and placing a stationary, rigid, conducting membrane between the
detector and the attractor. But this electrostatic shield places a practical
lower limit
of some tens of micrometers on the minimum attainable separation between the test bodies.

\subsubsection{Magnetic effects}
Microscopic particles of iron embedded in nominally nonmagnetic test bodies
during their machining or handling, or in the bulk during smelting, can create local
magnetic fields so small they
are difficult to detect with standard magnetometers, yet large enough to compete with
gravitational forces.
The magnetic force between two magnetically saturated iron particles $1$ mm apart,
each $10$ $\mu$m in diameter, can be as large as $10^{-7}$ dynes, varying as the
inverse
fourth power of the distance between the particles. This is as large as the
gravitational
attraction between a $1$-mm-thick Al plate with an area of $3$ cm$^2$ near an infinite
Al plate that
is $1$ mm thick. Yet the magnetic field of such a particle is only $0.3$ mGauss at a
distance of $2$ mm.

Most ISL tests modulate the position
of an attractor and detect the force this modulation produces on a detector.
Even if the attractor has no ferromagnetic impurities, any
magnetic field associated with the attractor modulation, e.g., from motor magnets or
flowing
currents, can couple to magnetic impurities in the detector. Experimenters typically
measure the magnetic field associated with the modulation of the attractor and apply
larger fields to
find the response of the detector.
A variety of smaller magnetic background effects are associated with the magnetic
susceptibilities
of the test bodies. Standard magnetic shielding of the experimental apparatus is usually
sufficient to
reduce the ambient magnetic field to a level where the susceptibilities pose no problem.
\subsubsection{Other effects}
Modulation of the attractor position may introduce
background effects that are not electromagnetic. The most obvious is a
spurious mechanical coupling that transmits the motion of the attractor through the apparatus to the detector.
These unwanted couplings can be reduced by multiple
levels of vibration isolation and by experimental designs that force the signal
frequency
to differ from
that of the attractor modulation. Experiments are performed in vacuum chambers to reduce
coupling between the test bodies from background gas.
\subsection{Experimental Strategies}
ISL tests can be constructed as null experiments,
partial-null  experiments, or direct measurements. For example, Hoskins et al.~\cite{ho:85}
studied the force on a cylinder located inside a cylindrical shell. To the
extent that the length-to-radius ratios of the cylinders are very large, this
constitutes a null test because the Newtonian interaction between the cylinders
gives no net force.
Other null tests have used planar geometry; the Newtonian force
between two parallel, infinite planes is independent of their separation.
This basic idea, as discussed below, was exploited in Reference~\cite{lo:02}.
An advantage of null experiments is that the apparatus does not need to handle
signals with a wide dynamic  range and the results are insensitive to
instrumental nonlinearities and calibration uncertainties.

Hoyle et al.~\cite{ho:01} have reported a partial-null experiment
in which the Newtonian signal was largely, but not completely, cancelled.
 As discussed below,
the partial
cancellation greatly reduced the required dynamic range of the instrument,
but Newtonian gravity still gave a very characteristic signal that was
used to confirm that the instrument was performing properly. The form and magnitude of this signal provided constraints on new physics.

Finally, Mitrofanov \& Ponomareva~\cite{mi:88} reported a direct experiment that compared the measured
force beween two spheres as their separation was switched between two values.
In this case, the results depended crucially on accurate measurement of the
separations of the spheres and the forces between them.
\section{EXPERIMENTAL RESULTS}
\subsection{Low-Frequency Torsion Oscillators}
\label{sec:low-f}
\subsubsection{The Washington experiment}
\begin{figure}
\vspace* {-0.2in}
\epsfysize=3.2in
\centerline{\epsfbox{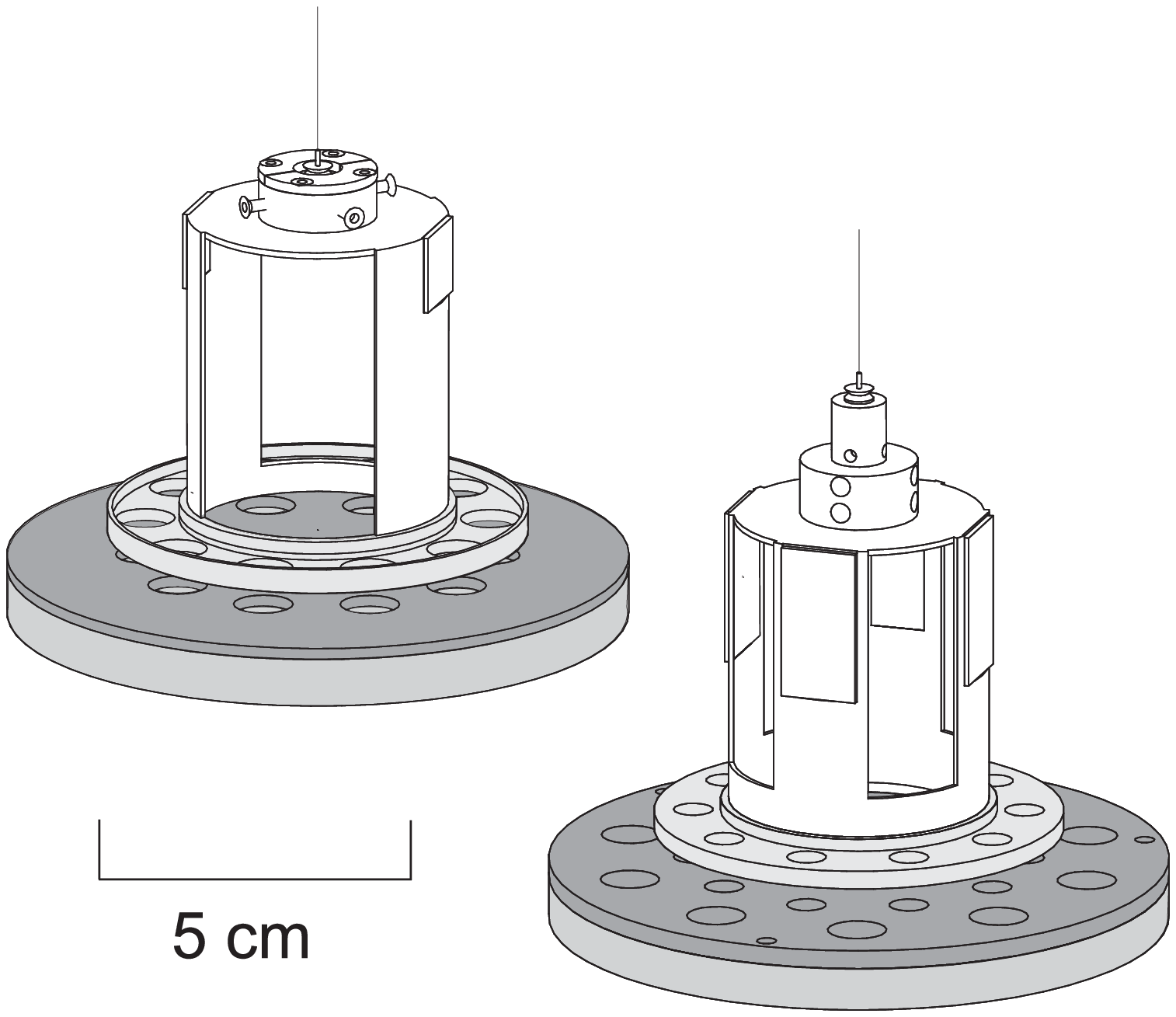}}
\caption{Schematic diagram of the 10-hole pendulums and rotating
attractors used in the two experiments of Hoyle et al. \cite{ho:01,ho:01a,he:02}. The active
components are shaded.}
\label{fig:hoyle apparatus}
\end{figure}

Hoyle et al. \cite{ho:01} of the University of Washington E\"ot-Wash
group developed a ``missing-mass''
torsion balance~(Figure \ref{fig:hoyle apparatus}),
for testing the ISL at short ranges. The active component of
the torsion pendulum was an
aluminum ring with 10 equally spaced
holes bored into it.
The pendulum was suspended above a copper
attractor disk containing
10 similar holes. The attractor was rotated uniformly by a geared-down
stepper motor.  The test bodies in this instrument were the ``missing''
masses of the two sets of
10 holes. In the absence of the holes, the disk's gravity
simply pulled directly down on the ring and did not
exert a twist. But because of the holes, the ring experienced a torque that
oscillated 10 times for every revolution of the disk---giving sinusoidal torques
at $10\omega$, $20\omega$, and
$30\omega$, where $\omega$ was the attractor rotation frequency.
This torque twisted the pendulum/suspension fiber and was
measured by an autocollimator that reflected a laser beam twice from a
plane mirror
mounted on the pendulum. Placing the signals at high multiples of the
disturbance frequency (the attractor rotation frequency) reduced many
potential systematic errors.
A tightly stretched 20-$\mu$m-thick beryllium-copper electrostatic shield
was interposed
between the pendulum and the attractor to minimize electrostatic and
molecular torques. The entire torsion  pendulum,
including the mirrors, was coated with gold and enclosed in a gold-coated
housing to minimize electrostatic effects. The pendulum could not ``see'' the
rotating attractor except for gravitational or magnetic couplings. Magnetic
couplings were minimized by machining
the pendulum and attractor with nonmagnetic tools and by careful handling.

\begin{figure}
\epsfysize=4.5in
\centerline{\epsfbox{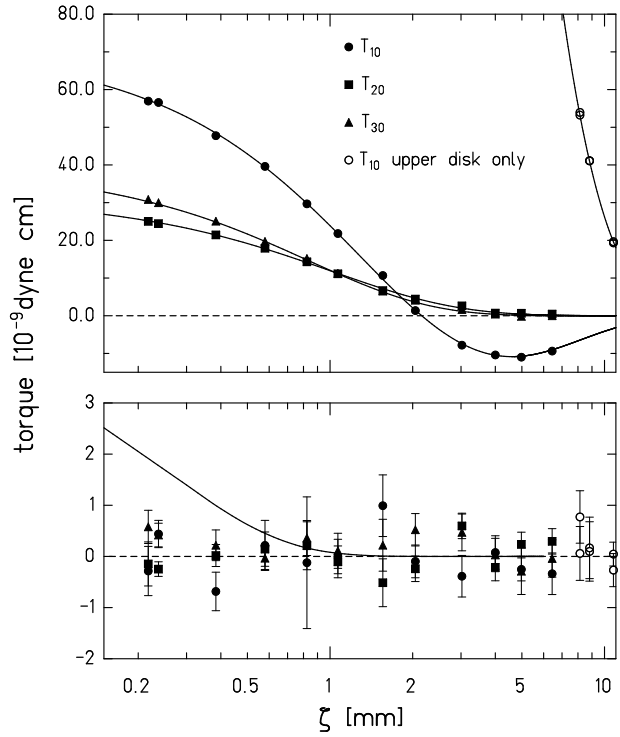}}
\caption{{\it Top:} Torques measured in the first experiment of Hoyle et al.
as a function of pendulum/attractor separation.
Open circles are data taken with the lower attractor disk removed and show the
effect of uncancelled gravity. Smooth curves show the Newtonian fit.
{\it Bottom:} Residuals for the Newtonian fit. The solid curve shows the
expected residual for a Yukawa force with $\alpha=3$ and $\lambda=250~\mu$m.}
\label{fig:hoyle torques}
\end{figure}

The experiment was turned into a partial-null measurement by adding a second,
thicker copper disk immediately below the upper attractor disk. This disk also had
10  holes, but they were rotated
azimuthally with respect to the upper holes by $18^\circ$ and their sizes
were chosen to give a $10\omega$ torque that
just cancelled the $10\omega$ Newtonian torque from the upper attractor.
On the other hand, a new short-range interaction would not
be cancelled because the lower attractor disk was simply too far  from
the pendulum. The cancellation was exact for a separation (between the lower
surface of the pendulum and the upper surface of the attractor) of about 2~mm.
For smaller separations the contribution of the lower disk was
too small to completely cancel the $10\omega$ signal, and at larger separations
the lower disk's contribution was too large (see Figure~\ref{fig:hoyle torques}).

Two slightly different instruments were used; both had
10-fold rotational symmetry and differed mainly in the dimensions of the
holes. In the first experiment, the pendulum ring was 2.002 mm thick with 9.545-mm-diameter
holes and a total hole ``mass'' of 3.972~g; in the second experiment, the ring
thickness was 2.979 mm  with 6.375-mm-diameter holes having a total hole
``mass'' of 2.662 g. The resonant frequencies of the two  pendulums, $\omega_0/2\pi$, were 2.50 mHz and 2.14 mHz, respectively; the fundamental $10 \omega$ signals were set at
precisely $\textstyle{\frac{10}{17}}$ $\omega_0$ and $\textstyle{\frac{2}{3}}$ $\omega_0$, respectively. In both cases, the
$20\omega$ and
$30\omega$ harmonics were above the resonance.
The observed spectral density of deflection noise was close to the thermal value
given in Equation~\ref{eq:thermal} for the observed $Q$ factor of 1500 (see also
Figure~\ref{fig:22 hole FFT} below).
%
%
%
\subsubsection{Signal scaling relations}
\label{sec:low-f signal}
The gravitational torque exerted on the pendulum by the rotating attractor is
$T_g(\phi)=-\partial V(\phi)/\partial \theta$, where $V(\phi)$ is the
gravitational potential energy of the attractor when the attractor is at angle $\phi$,
and $\theta$ is the twist angle of the pendulum. For cylindrical holes, four
of the six Newtonian torque integrals can be solved analytically but the remaining two
must be evaluated numerically. Clearly, the Newtonian signal drops as the number of holes
increases and their radii decrease because the long-range gravitational force tends
to ``average away'' the holes. It also drops rapidly for
separations much greater than the thickness of the upper attractor disk.
Only three of the Yukawa torque integrals can be solved analytically.
However, when the Yukawa range, $\lambda$,
becomes much smaller than any of the relevant dimensions of
the pendulum/attractor system, a simple scaling relation based on
Equation~\ref{eq:plate yukawa potential} governs the signal and
\begin{equation}
T_Y \propto \alpha G \rho_p \rho_a \lambda^3 e^{-s/\lambda} \frac{\partial A}{\partial
\phi},
\label{eq:scaling}
\end{equation}
where $\rho_p$ and $\rho_a$ are the densities of the pendulum and  attractor, respectively; $\lambda$
is
the Yukawa  range; and $A$ is
the overlap area of the holes in the pendulum with those of
the attractor when the attractor angle is $\phi$.
\subsubsection{Backgrounds}
Hoyle et al.~\cite{ho:01} found that the effects from spurious gravitational couplings, temperature fluctuations, variations
in the tilt of the
apparatus, and magnetic couplings were  negligible compared  with
the statistical errors. 
Electrostatic couplings were negligible because the pendulum was almost completely
enclosed
by a gold-coated housing. The 20-$\mu$m-thick electrostatic shield was rigid to prevent
secondary electrostatic couplings.
The shield's lowest resonance was about 1 kHz, and the attractor could  produce a
false electrostatic effect only by flexing the shield at a very high $m=10$ mode.
\subsubsection{Alignment and calibration}
Although all submillimeter tests of the ISL face an alignment problem,
it was especially important in this experiment because of the
relatively large size of the pendulum (chosen to increase the
sensitivity). Alignment was done in stages. First the pendulum ring was
leveled by nulling its differential capacitance as the pendulum
rotated above two plates installed in place of the electrostatic
shield. The shield was then replaced, and the tilt of the entire apparatus
was adjusted to minimize the pendulum-to-shield capacitance.
To achieve horizontal alignment, the gravitational torque was measured 
as the horizontal position of the upper
fiber suspension point was varied. Determining separations
from mechanical or electrical contacts gave unreliable results,
so the crucial separation between the
pendulum and the electrostatic shield was determined from the
electrical capacitance.

The experimenters calculated the torque scale directly, using gravity.
Two small aluminum spheres
were placed in an opposing pair of the 10 holes of the torsion pendulum
and two large bronze spheres, placed on an external turntable, were rotated uniformly
around the instrument at a radius of 13.98 cm. Because this was close
to the 16.76-cm radius \cite{gu:00} used in determining $G$ and
the ISL has been tested at this length scale
(see Figure~\ref{fig:alpha lambda long}),
the calibration torque could be computed to high accuracy.
The torsion constant of the fiber was about 0.03 dyne cm. 

\subsubsection{Results}
\label{sec:ew results}
Data were taken at pendulum/attractor separations down to 197 $\mu$m,
where the minimum separation was limited by pendulum ``bounce'' from seismic
disturbances.
The torque data, shown in Figure~\ref{fig:hoyle torques}, were analyzed by fitting
a potential of the form given in Equation~\ref{eq:yukawa} with
$\alpha$ and $\lambda$ as free parameters
and treating the important experimental parameters (hole masses and dimensions,
zero of the separation scale, torque calibration constant, etc.)
as adjustable
parameters constrained by their independently measured values.
Hoyle et al.\ reported results
from the first of the two experiments in Reference \cite{ho:01}; 
the combined 95\%-confidence-level (CL)
result of both experiments
was given  subsequently \cite{ho:01a,he:02} and is shown in
Figure~\ref{fig:alpha lambda}.

The results exclude the scenario of two equal extra dimensions
whose size gives a unification scale of $M^{\ast}=1$~TeV; this would
imply an effective Yukawa interaction with $\lambda= 0.3$ mm and $\alpha=16/3$
if the extra dimensions are compactified as a
torus. Because $\alpha \ge 16/3$ is consistent with the data only
for $\lambda < 130~\mu$m, Equation~\ref{add} implies that $M_{\ast} > 1.7$ TeV.
A tighter bound on $M_*$ can be extracted from the radion constraint,
which, in the unwarped case where $1/3 \leq \alpha \leq 3/4$ for 
$1 \leq n \leq 6$, suggests that $M_* \geq \CO(3~{\rm Tev})$.

More interesting and general is the upper limit 
placed on the size of the 
largest single
extra dimension, assuming all other extra dimensions are significantly smaller \cite{ho:01,he:02}.
For toroidal compactification, this corresponds to the largest
$\lambda$ consistent with $\alpha=8/3$, leading to an upper limit 
$R_* \leq 155~\mu$m. Other compactification schemes necessarily give
somewhat different limits.

\subsection{High-Frequency Torsion Oscillators}
\label{sec:high-f}
\subsubsection{The Colorado experiment}
The modern era of short-range ISL tests was initiated by Long et al.\
at the University of Colorado \cite{lo:99}. Their apparatus, shown in
Figure~\ref{fig:long apparatus}, used a planar null geometry.
The attractor was a small 35 mm $\times$ 7 mm $\times$ 0.305 mm tungsten
``diving board'' that was driven vertically at 1 kHz in its  second
cantilever mode by a
PZT (lead zirconate titanate) bimorph. The detector, situated below the diving board, was an unusual
high-frequency compound torsion oscillator made from 0.195-mm-thick tungsten.
It consisted of a double rectangle for which the  fifth normal mode resonates
at 1 kHz; in this mode, the smaller 11.455 mm $\times$ 5.080 mm rectangle
(the detector) and the larger rectangle (one end of which was connected to a
detector mount)
counter-rotated about the torsional
axis, with the detector rectangle having the larger amplitude. The
torsion oscillations were read out capacitively from the larger rectangle.
The attractor
was positioned so that its front end was aligned with the back edge of the
detector rectangle and a long edge of the attractor was aligned above
the detector torsion axis. A small electrostatic shield consisting of a  0.06-mm-thick
sapphire plate coated with 100 nm of gold was suspended between the attractor
and the detector. The attractor, detector, and electrostatic shield were mounted on
separate vibration-isolation stacks to minimize any mechanical couplings and
were aligned by displacing the elements and measuring the points of mechanical contact.

In any null experiment, it is helpful to know the precise form of a
signal of new physics. Long et al. slid away the electrostatic shield and applied a 1.5-V bias to the
detector to give a large, attractive electrostatic force; this determined
the phase of the signal that would be produced by a new, short-range
interaction.
\subsubsection{Signal-to-noise considerations and calibration}
The spectral density of thermal-force noise in the multimode oscillator
used in Reference~\cite{lo:99} obeys a relation
similar to Equation~\ref{eq:torque noise}. The Colorado
experimenters operated on a resonance
with a $Q=25,000$ so the readout noise was negligible.
Data were taken with the attractor driven at the detector resonance as well
as about 2 Hz below the resonance (see Figure~\ref{fig:long data}).
The mean values of the on-resonance and
off-resonance data agreed within errors, but the standard deviation of
the on-resonance data was about twice that of the off-resonance data.
This is just what one would expect if the on-resonance data were dominated by
thermal noise. Furthermore, the on-resonance signal did not change
as the geometry was varied. This ruled out the unlikely possibility
that the observed null result came from a fortuitous cancellation of
different effects, all of which should have different dependences
on the geometry. The torsion oscillation scale was calibrated
by assuming that the on-resonance signal was predominantly thermal.
\subsubsection{Backgrounds}
Although a net signal was seen, it had the same magnitude on and
off resonance and presumably was due to electronic pickup.
No evidence was seen for an additional, statistically significant background.
Checks with exaggerated electostatic and magnetic effects showed
that plausible electrostatic and magnetic couplings were well
below the level of thermal noise.
\subsubsection{Results}
The null results from this experiment, taken at a separation of 108~$\mu$m,
were turned into $\alpha(\lambda)$ constraints using a maximum-likelihood technique.
For various assumed values of $\lambda$, the expected Yukawa force was calculated
numerically
400 times, each calculation using different values for experimental parameters that were
allowed to vary within their measured ranges. A likelihood function  constructed
from these calculations  was used to extract 95\%-CL limits on $\alpha(\lambda)$.
The results \cite{lo:02}, shown in Figure~\ref{fig:alpha lambda},
exclude a significant portion of the moduli forces predicted
by Dimopoulos \& Giudice \cite{DG:96}.
\subsection{Microcantilevers}
\label{sec:micro}
\subsubsection{The Stanford experiment}
Chiaverini et al.\ at Stanford \cite{ch:02,ch:02a} recently reported a test of the
ISL using the microcantilever apparatus shown schematically in Figure~\ref{fig:kapitulnik instrument}. This instrument was suited for the 10-$\mu$m
length  scale but lacked the
sensitivity to see gravity. The apparatus consisted of a silicon microcantilever with
a 50~$\mu$m $\times 50$~$\mu$m $\times 50$~$\mu$m gold test mass mounted
on its free end.
The cantilever had a spring constant of about 5 dyne/cm,
and its displacement was read out with an optical-fiber interferometer.
The microcantilever, which hung from a two-stage
vibration-isolation system, oscillated vertically in  its lowest flexural mode
at a resonant frequency of $\omega_0 \approx 300$~Hz . The
microcantilever was mounted above an attractor consisting of  five pairs of alternating
100~$\mu$m $\times 100$~$\mu$m $\times 1$~mm bars of gold and silicon.
The attractor was oscillated horizontally underneath the cantilever at about
100 Hz by a bimorph; the amplitude was chosen to effectively resonantly excite the
cantilever
at the  third harmonic of the attractor drive frequency.
The geometry was quite
complicated; the  third harmonic gravitational force on the cantilever depended
sensitively and nonlinearly on the drive amplitude.
An electrostatic shield consisting of a 3.0-$\mu$m-thick silicon nitride
plate with 200 nm of gold evaporated onto each side was placed between the
cantilever and the attractor.
Data were taken with the vertical separations between the cantilever
and the attractor as small as $25~\mu$m.
\subsubsection{Signal-to-noise considerations}
The dominant noise source in the Stanford experiment was thermal noise in
the cantilever, which was reduced by operating at about 10 K.
The  $Q$ factors of the oscillating cantilevers in these measurements
were typically about 1200.
\subsubsection{Calibration and alignment}
The cantilever spring constant $k$ was found in two independent ways that agreed
to within 10\%: by assuming
that when the cantilever was far from the attractor it was in thermal
equilibrium with its surroundings, and by calculating $k$ from the
measured resonant frequency. The
cantilever was aligned with respect to the attractor using magnetic forces.
The cantilever's test mass had a thin nickel film on one  face,
and the attractor was equipped with a zig-zag conducting path  that followed the
gold bars. When a current was run through the attractor, it placed a force
on the cantilever that had half the frequency and phase  of the
expected gravitational signal but  vastly greater amplitude. This
force was used to align the apparatus.
\subsubsection{Backgrounds}
This experiment was limited by a spurious force about 10 times greater than the
thermal detection limit. This force was clearly not fundamental,
i.e., related to
the mass distributions on the attractor, because the
phase of signal did not behave as expected when the horizontal offset
of the attractor oscillation was varied or as the attractor drive amplitude
was changed. The most likely source of a spurious force is electrostatics;
the cantilever was not metallized and so it could hold charge and the shield was
observed to vibrate by a picometer or so. A potential on the cantilever of about
1 V would be sufficient to produce the observed force.
Although thin nickel layers were
incorporated into the
test mass and attractor, the experimenters estimate that magnetic forces
from the nickel (as well as from iron impurities in the gold) were too small
to explain the observed  background force.
Vibrational coupling
between the attractor and cantilever was minimized because the attractor
was moved at right angles to the cantilever's flex.
\subsubsection{Results}
The experimenters saw a spurious $(8.4 \pm 1.4) \times 10^{-12}$ dyne force when the attractor and cantilever were at their
closest separation of $25\mu$m. They assigned a 95\%-CL upper limit on
a Yukawa interaction by computing the minimum
$\alpha$ as a function of $\lambda$ that would correspond to this
central value plus two standard deviations. Figure~\ref{fig:alpha lambda short} shows their constraint, which rules out
much of the parameter space expected from moduli exchange as computed
in Reference \cite{DG:96}.
\subsection{Casimir Force Experiments}
\label{sec:casimir experiments}
Early attempts to detect the Casimir force between metal surfaces \cite{sp:58} and
dielectric
surfaces \cite{de:56,ta:69,is:72,hu:72} had relatively large errors.
Nonetheless, it was recognized \cite{ku:82,mo:87,efi:92}
that such measurements provided the tightest constraints on new
hypothetical particles with Compton wavelengths less than $0.1$ mm. In recent years,
three groups have reported measurements of the Casimir force with relative errors of
$1\%$ to $5\%$.
Although these experiments are orders of magnitude away from providing tests of the ISL,
they do probe
length scales from 20 nm to 10 $\mu$m,
where large effects may occur (see Section~\ref{sec:hidden}).
\subsubsection{Experimental methods}
The first of the recent experiments, performed by Lamoreaux at the University of
Washington \cite{la:97,lam:98}, used a torsion balance
to measure the force between a flat quartz plate and a spherical lens with a radius
of $12.5 \pm 0.3$ cm. Both surfaces were coated with 0.5~$\mu$m of copper followed by 0.5~$\mu$m of gold.
A piezoelectric stack stepped the separation between the plate and lens from 12.3~$\mu$m
to
0.6~$\mu$m, at which point the servo system that held the torsion pendulum angle
constant became
unstable. The force scale was calibrated to $1\%$ accuracy by measuring the servo
response when
a $300$-mV potential difference was applied between the plate
and lens at a large ($\approx 10$~$\mu$m) separation.
The absolute separation between the lens and plate was obtained by applying a potential
difference between the two surfaces and fitting the measured force (for distances
greater than
2~$\mu$m where the Casimir force was small) to the expected $1/d$ dependence, where $d$
is the
distance between the plate and lens. After subtracting the $1/d$ component from the
force scans,
the residual signals were fitted to the expected form for a Casimir force, and they agreed to within $5\%$ \cite{la:97,lam:98}.

Mohideen and collaborators at the University of  California at Riverside reported a series
of experiments that used an atomic-force microscope (AFM) to measure the Casimir force
between a small sphere and a flat plate \cite{umo:98,roy:99,aro:99,har:20}.
Their most recent measurement used a
191-$\mu$m-diameter polystyrene sphere that was glued to a 320-$\mu$m-long AFM
cantilever.
The cantilever plus sphere and a $1$-cm-diameter optically polished sapphire disk were coated with $87$ nm of  gold, with a
measured surface
roughness of $1.0 \pm 0.1$ nm. The disk was placed on a piezoelectric tube with the
sphere
mounted above it, as shown in Figure~\ref{fig:afm}.
The cantilever flex was measured
by reflecting laser light from the cantilever onto split photodiodes. The force scale
was
calibrated electrostatically by applying a $\pm 3$-V potential difference
between the sphere and disk at a separation of 3~$\mu$m.
The force difference between the $+3$ V and $-3$ V applied potentials was used
to determine the residual potential difference between the disk and sphere when their
external
leads
were grounded together: $3 \pm 3$ mV. The force between the sphere and disk was measured
for
separations ranging from $400$ nm to contact.
It was found that the surfaces touched when their average
separation was $32.7 \pm 0.8$ nm. This was attributed to  gold crystals protruding from the
surfaces. The
measured forces were compared to the expected Casimir force for separations  of
62--350 nm and agreement to within $1\%$ was found \cite{har:20}.

The record for measuring the Casimir force at the closest separation is held by
Ederth \cite{ede:20} at the Royal Institute of Technology in Stockholm, who
measured the force between crossed cylindrical
silica disks with diameters of $20$ mm.
A template-stripping method \cite{wag:95} was used to glue $200$-nm layers of  gold, with an
rms surface
roughness of $\le 0.4$ nm, to the silica disks. The  gold surfaces were then coated with a
$2.1$-nm-thick
layer of hydrocarbon chains to prevent the adsorption of surface contaminants and the
cold-welding of the  gold surfaces upon contact.
One cylindrical surface was attached to a piezoelectric
stack and the other to a piezoelectric bimorph deflection sensor that acted as a
cantilever spring.
The two surfaces were moved toward one another starting at a separation $>1$~$\mu$m, where the Casimir force was less than the  resolution of the force sensor,
and ending at a separation of $20$ nm, at which point
the gradient of the
Casimir force was comparable to the stiffness of the bimorph spring, causing the
surfaces to
jump into contact. The stiffness of the bimorph sensor was calibrated by continuing to
move the piezotube
another 200--300 nm while the surfaces were in contact. The absolute separation between
the surfaces was
found by fitting the measured force curve to the expected Casimir signal
(plus electrostatic background, which was found to be negligible) with the absolute
separation as
a fit parameter. At contact, the surfaces compressed by $\approx 10$
nm.
The measured force  was compared to  the expected Casimir force
over the range of separations from $20$ to $100$ nm  and an agreement  to better than $1\%$ was found.
\subsubsection{Signal-to-noise and background considerations}
The signal-to-noise ratio for Casimir-force measurements as tests of the ISL
may be improved by using more sensitive force probes, using thicker metallic coatings on the
test bodies,
and operating at lower temperatures. Nonetheless, the dominant limitation
for interpreting the measurements as tests of the ISL   is understanding the
Casimir-force background
to high accuracy.
There is a growing literature on the corrections that must be applied to the Casimir
force calculated
for smooth, perfect conductors at zero temperature (Equations~\ref{eq:cas1} and~\ref{eq:cas2}). The
dominant corrections are for finite temperature, finite conductivity, and surface
roughness.
Corrections for finite temperature are important for test-body separations 
$d > 1$~$\mu$m.
%
%
%
%
For the Lamoreaux experiment, the finite-temperature corrections at 1-$\mu$m
and 6-$\mu$m separations were $2.7\%$  and $174\%$
of the zero-temperature Casimir force, respectively \cite{bor:98}. A number of authors have considered the effects of finite
conductivity on the temperature correction  \cite{kli:01,gen:00,bord:00}, and results
believed to be accurate to better than $1\%$ were obtained.
The correction to the Casimir force for the finite conductivity of the metallic surfaces
is of order
$10\%$ at $d =1$~$\mu$m and grows with smaller separations. Finite-conductivity
corrections using a plasma model for the dielectric function of the metal give the
correction as a power
series in $\lambda_P/d$, where $\lambda_P$ is the plasma wavelength of the
metal \cite{sch:78,bez:97,klim:99}.
Corrections have also been obtained using optical data for the complex dielectric
function \cite{lamo:99,har:20,lamb:00,bos:00}. Surface roughness of the test bodies
contributes a correction to the Casimir force that can be expressed as a power series in
$h/d$, where
$h$ is a characteristic amplitude of the surface
distortion \cite{vb:74,mara:80,klim:96,klim:99}.
For stochastic distortions, the leading-order surface-roughness correction is $6(h/d)^2$,
which is less
than $1\%$ of the Casimir force at closest separation in the experiments of Ederth and
the Riverside
group.
\subsubsection{Results}
\label{sec:casimir results}
Constraints on Yukawa interactions with ranges between $1$ nm and
$10 \mu$m, shown in 
Figure~\ref{fig:alpha lambda short},
have been extracted from the Casimir-force measurements of
Lamoreaux \cite{bor:98,lo:99},
Ederth \cite{most:01}, and
the Riverside group \cite{bord:99,bor:00,fi:01,mos:00}.
Figure~\ref{fig:alpha lambda short} also shows constraints at even
smaller ranges obtained from earlier van der Waals--force experiments \cite{bor:94}.
It should be noted that most of these constraints were obtained by assuming that a 
Yukawa force could not exceed the difference between the measured force and
the predicted Casimir effect. To be rigorous, the raw data should be
fitted simultaneously with both Casimir and Yukawa forces, 
which should lead to significantly less stringent limits on $|\alpha|$.
Deviations from Newtonian gravity in this region that follow a power law 
(Equation~\ref{eq:power law}) are  constrained  more
strongly by the much more sensitive longer-range gravity experiments
discussed above \cite{fi:01}.
\subsection{Astronomical Tests}
\label{sec:astro}
A summary of constraints on Yukawa interactions with $\lambda \ge 1$~mm
may be found in Figure 2.13 of the 1999
review by Fischbach  \& Talmadge \cite{fi:99}, which we reproduce
in part in our Figure~\ref{fig:alpha lambda long}. Since the publication of Reference~\cite{fi:99}, 
the constraints for $\lambda \leq 1$~cm have been
substantially improved, as discussed above. The constraints at
larger ranges from
laboratory, geophysical, and astronomical data (see Figure~\ref{fig:alpha lambda long})
are essentially unchanged. The astronomical
tests provide the tightest constraints on $\alpha$. These are typically based
on Keplerian tests comparing $G(r) M_{\odot}$ values deduced for different planets. However,
the tightest constraint comes from lunar-laser-ranging (LLR) studies of
the lunar orbit. Because this result may improve significantly in the next
few years, we give some details of the measurement here.

The LLR data consist of range measurements from telescopes on
Earth to retroreflectors placed on the Moon by US astronauts
and an unmanned Soviet lander. The measurements, which began in
1969, now have individual raw range precisions of about 2 cm
and are obtained from single photon returns, one of which is detected
for roughly every 100 launched laser pulses \cite{llrgeneral}.
The vast majority of
the data come from sites in Texas \cite{sh:85} and in southern
France \cite{sa:98}. The launched laser pulses have full widths
at half maximum of about 100 ps; the return pulses are broadened
to about 400 ps because the reflector arrays typically do
not point straight back to  Earth owing to lunar librations.
The  launch-telescope--to--lunar-retroreflector ranges
have to be corrected for atmospheric delay,
which is computed from the local barometric pressure, temperature,
and humidity.
For the Moon straight overhead, the range correction at the
Texas site is about 2 m. The dominant uncertainties in converting
raw range measurements into
separations between the centers of mass of the
Earth and the Moon
come from tidal distortions of the Earth and Moon and atmospheric
and ocean loading of the Earth.
The current model, using the entire world data set, gives an uncertainty
of about 0.4 cm in the important orbit parameters.

The most sensitive observable for testing the ISL
is the anomalous precession of the lunar orbit.
If the Moon were subject only to a central Newtonian $1/r$ potential from the
Earth, the lunar orbit would not precess. The orbit does precess due to the Earth's 
quadrupole field and perturbations from other solar-system bodies, as well
as from the small
general relativistic geodetic precession and possibly also from
a Yukawa interaction; the conventional sources
of precession must be accounted for to obtain the anomalous
Yukawa precession
rate. Ignoring terms of order $\varepsilon^2$, where the Moon's eccentricity
is $\varepsilon=0.0549$, the anomalous Yukawa precession rate $\delta\omega$
is \cite{fi:99}
\begin{equation}
\frac{\delta\omega}{\omega} = \frac{\alpha}{2} \left( \frac{a}{\lambda}
\right)^2 e^{-a/\lambda}~,
\end{equation}
where $\omega=2\pi$ radians/month and $a$ is the mean radius of the Moon's orbit.
The constraint on $\alpha(\lambda)$ is tightest for $\lambda=a/2$
and falls off relatively steeply on either side of $\lambda=a/2$.
The current LLR $2\sigma$ upper limit on $\delta\omega$ is 270 $\mu$as/y; this
follows because
the observed precession of about 19.2  mas/y 
agrees with the general relativistic prediction to
$(-0.26 \pm 0.70)\%$, where the error is ``realistic'' rather than
``formal'' (the error quoted in Reference~\cite{wi:02} should be doubled;
J.\ Williams,
private communication 2003). We conclude that at 95\%  CL,
$\delta\omega/\omega < 1.6\times10^{-11}$; the corresponding LLR
constraint is shown in Figure~\ref{fig:alpha lambda long}.
\section{CONCLUSIONS}
\subsection{Summary of Experimental Results}
\label{sec:summary}
Because gravity is intimately connected to the geometry of spacetime,
ISL tests could provide very direct evidence for the existence of extra
space dimensions. In addition, ISL tests are sensitive to the
exchange of proposed new low-mass bosons. A variety of theoretical
considerations hint that new effects may occur at
length scales between
10 $\mu$m and 1 mm. This circumstance, as well as the urge to explore
unmapped territory, has motivated the development of new experimental
techniques that have produced substantial improvements in constraints
on theories.
The overall slope of the experimental constraints
shown in Figures~\ref{fig:alpha lambda}, \ref{fig:alpha lambda short}, and
\ref{fig:alpha lambda long} reflects the rapidly decreasing signal strength
of a new interaction as its range decreases.
At gravitational strength ($\alpha=1$ in Figure~\ref{fig:alpha lambda}),
the ISL has been verified down to a distance $\lambda=200~\mu$m. At length
scales between 20 nm and 4 mm, many square decades in Yukawa-parameter space
have been ruled out. These results have eliminated some specific
theoretical scenarios, but many other interesting ideas are still viable  because
their predicted effects lie somewhat below the current experimental limits.
\subsection{Prospects for Improvements}
\subsubsection{Short-range tests of the ISL}
To make a gravitational-strength ($\alpha=1$) ISL test at a 20-$\mu$m length scale requires an
increase in the background-free sensitivity of at least a factor of
$10^3$. Fortunately, such an increase is possible, although it will
require years of development.

The E\"ot-Wash group are currently running a new apparatus that features
a pendulum/attractor system having 22-fold rotational
symmetry with 44 thinner, smaller-diameter holes. The pendulum ring and
attractor disk are made from denser materials (copper and molybdenum,
respectively). Noise has been improved by a factor of six. The closest
attainable separation has been reduced by a factor of two by adding
a passive ``bounce''-mode damper to the fiber-suspension system, and
the thickness of the electrostatic shield has been reduced to
10 $\mu$m. Figure~\ref{fig:22 hole FFT}
shows the spectral density of the torque signal from this apparatus.
This instrument should probe Yukawa forces with $|\alpha|=1$ for ranges
down to $\lambda=60~\mu$m. In principle, it is possible to use
a low-frequency torsion balance in a different mode, one that measures
the attraction between two flat plates (J.G.\ Gundlach, private communication). This would provide a null test with
a sensitivity that scaled as $\lambda^2 e^{-s/\lambda}$ rather than as
$\lambda^3 e^{-s/\lambda}$ in the partial-null experiments.

The Colorado group plans to optimize their geometry and to use a
Washington-style electrostatic shield to attain closer separations. This
could improve their limits between $10~\mu$m and $50~\mu$m by at least
an order of magnitude.
In the long run, both groups  could run at liquid-helium temperatures, which will
give lower noise, not only from the decreased $k_BT$ factor, but also from the
expected increase in the  $Q$ factor  of the torsion oscillator. Newman \cite{ba:00}
found that the  $Q$ factor of a torsion fiber has two  components.
One  is temperature-independent but amplitude-dependent (this is
already negligible in the E\"ot-Wash instrument because of the small amplitudes
employed) and the other is temperature-dependent and amplitude-independent.

The microcantilever application exploited by the Stanford group has
not yet attained its full potential. Presumably, lessons learned in
this pioneering experiment will reduce the backgrounds and allow the
experimenters to exploit their inherent sensitivity to new
very small forces.
Because corrections to the idealized Casimir force can be large and depend on
properties of the test
bodies that are troublesome to quantify, it may be difficult to compare Casimir-force experiments to theory
at an accuracy much better than $1\%$. The finite-conductivity corrections depend on
the dielectric
properties of the actual metallic coating of the test  bodies, which may differ somewhat
from bulk
dielectric properties used in the calculation. As the experimental precision improves,
parameters
associated with the conductivity correction (such as $\lambda_P$) may need to be
included as adjustable parameters in fitting the measured  force-versus-distance  curves. 
The surface-roughness correction should consider
distortions over  length  scales larger than are easily accessible by AFM scans, and it
may be necessary to vary the roughness parameters as well.
Both corrections scale as inverse
powers of the separation, $d$, as do the corrections for residual electric potential and
patch
effects.
Compounding the problem of multiple
corrections with similar distance dependences is the
uncertainty in the absolute separation of the test bodies. The Casimir force
depends on $d_0 + d_r$ rather than on $d$, where $d_r$ is the relative displacement
of the test bodies between force measurements (which can be accurately measured) and
$d_0$ is the absolute
separation at the origin of the relative scale (which is difficult to determine
accurately).
Including $d_0$ as a fit parameter allows
other short-distance parameters to vary \cite{ede:20}
without affecting the fit at large distances, where the
fractional error on the force measurements is larger. It is unlikely that the next few
years will  bring large improvements in Yukawa constraints from Casimir-force experiments.
\subsubsection{Long-range tests of the ISL}
Because any change in orientation of the Moon's
ellipticity grows linearly with time, even with data of
constant precision the LLR constraint should improve in proportion
as the data span increases (assuming that the modeling of conventional
precession sources is not a limiting factor).
New LLR projects should improve the raw range precision by an
order of magnitude, bringing the precision into the range needed to test the 
``high-pass'' gravity model \cite{newref_giu02}. For example, APOLLO \cite{mu:00} will
exploit a 3.5-m telescope at an elevation of 2780 m and subarcsecond image
quality. This instrument
should receive several returned photons per laser shot, giving a data
rate about $10^3$ times greater than existing facilities. It is
expected that more precise data will lead to corresponding improvements
in the modeling.

Ranging to other   planets is necessary in order to
probe longer length scales  effectively. This is currently done using radar
(which is limited by the absence of a well-defined ``target'' on the
planet) or  else microwave signals transmitted
by orbiting spacecraft (which are limited by uncertainties and the finite
timespan of the orbits). Furthermore, the accuracy of 
microwave ranges is
limited by propagation delay in the interplanetary solar plasma.
It is impractical to laser-range to passive reflectors
on other planets (if they could be placed) because the returned signal
falls as $1/r^4$. However, recent developments in active laser transponders,
whose sensitivity falls as $1/r^2$,
make it practical to place such a device on Mars and ultimately achieve
range precisions of a few centimeters \cite{de:02}. This would yield
several interesting new gravitational measurements, including an improved
test of the strong equivalence principle \cite{no:}, which provides one of
the best limits on massless gravitational scalar fields,
as well as tests of the ISL that would give interesting constraints on
the quasilocalized gravity model of Reference~\cite{Dv:02a}.

ISL tests at scales larger than the solar system typically rely
on uncertain astrophysical models. But Will \cite{wi:98} notes that
the proposed LISA space-based interferometer could
test a pure Yukawa potential at a scale of $5 \times 10^{19}$ m
by studying distortions of the gravitational waveform from an
inspiraling pair of
$10^6 M_{\odot}$ compact objects.

\subsection{What if a Violation of the $1/r^2$ Law Were Observed?}
Suppose that future experiments revealed a violation of the ISL
at short length scales.
Of course, one would try to tighten the constraints on its range and strength by
performing tests using instruments with varying length scales. But
a new question immediately arises: Is the new physics a geometrical effect of
extra dimensions or evidence for exchange of a new boson? This can
be decided by testing whether the short-range interaction violates the
equivalence principle: Boson exchange
generically does not couple to matter in a universal manner and therefore
appears as a ``violation'' of the equivalence principle, whereas geometrical
effects must respect the principle. Kaplan \& Wise~\cite{Ka:00a} estimated that the
equivalence-principle-``violating'' effect from dilaton exchange
is $\approx 0.3\%$.
\section{ACKNOWLEDGMENTS}
This work was supported in part by the National Science Foundation (Grant PHY-997097)
and by
the Department of Energy. We are grateful for discussions with Z.\ Chacko, K.\ Dienes,
G.\ Dvali, S.\ Dimopoulos, J.\ Erlich, P.\ Fox, G.\ Giudice, J.\ Gundlach, N.\ Kaloper,
E.\ Katz, T.\ Murphy, R.\ Rattazzi, and J.\ Williams.
D.B.\ Kaplan collaborated significantly on Sections 2.41--2.44.
D.\ Kapner and E.\ Swanson helped with the figures.

\section{ LITERATURE CITED}

\newpage

%
%
\begin{figure}
\epsfysize=4.2in
\centerline{\epsfbox{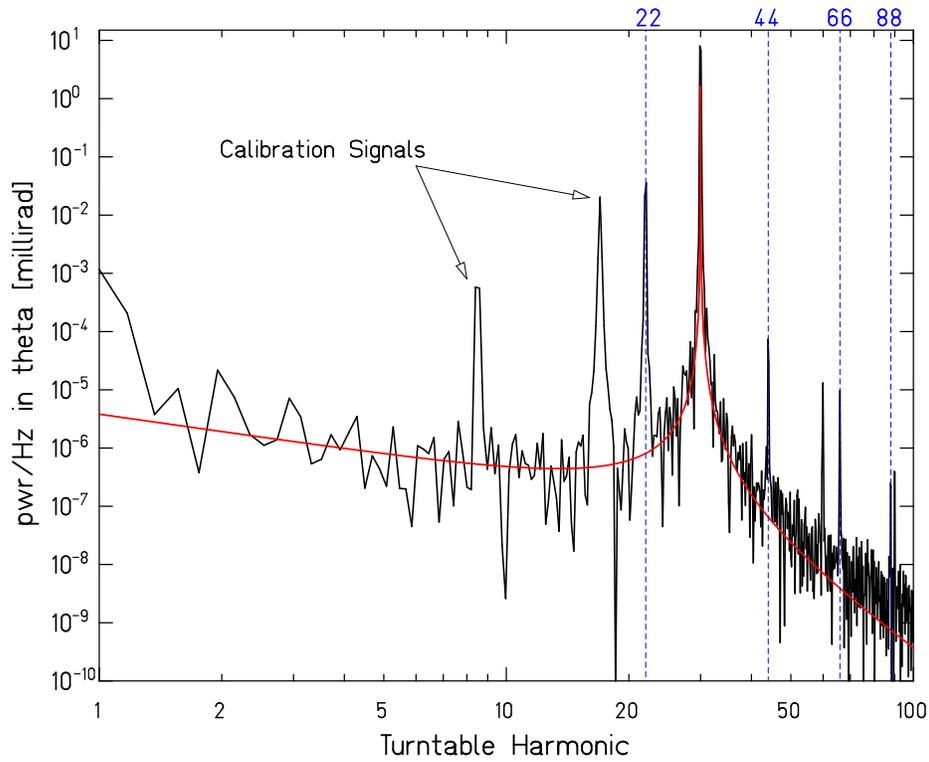}}
\caption{Spectral density of the torque signal in the 22-fold symmetric experiment
of the E\"ot-Wash group. The peaks at 8.5 and 17 $\omega$ are gravitational
calibrations; the fundamental and first three overtones of the short-range signal
are at 22, 44, 66, and 88 $\omega$. 
The smooth curve shows the thermal noise computed using Equation~\ref{eq:thermal}.}
\label{fig:22 hole FFT}
\end{figure}

\begin{figure}
\epsfysize=4.2in
\centerline{\epsfbox{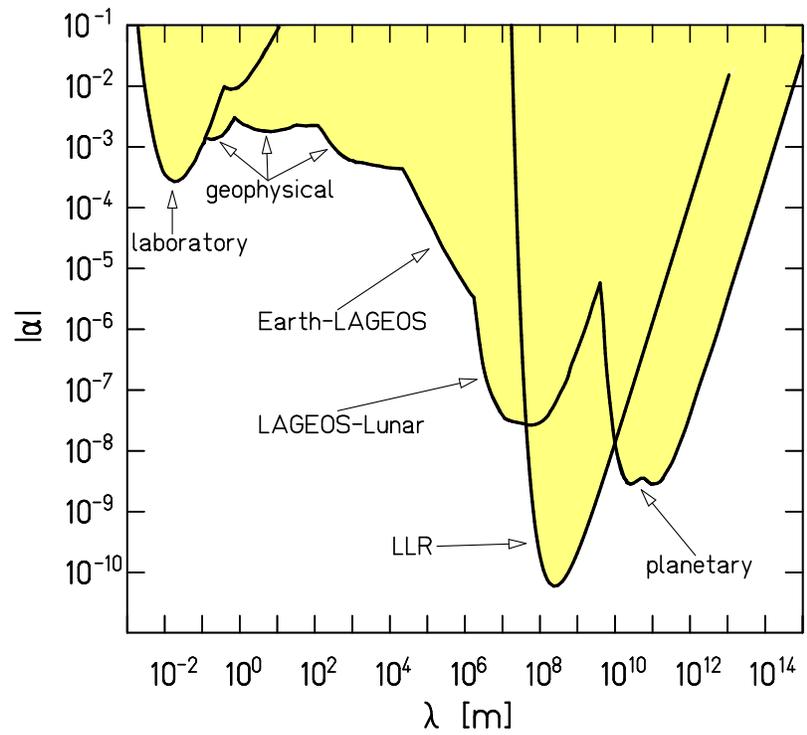}}
\caption{95\%-confidence-level constraints on ISL-violating Yukawa
interactions with $\lambda > 1$ cm. The LLR constraint is based
on the anomalous
perigee precession; the remaining constraints are based on Keplerian
tests. This plot is based on Figure 2.13 of
Reference~\cite{fi:99} and updated to include recent LLR results.}
\label{fig:alpha lambda long}
\end{figure}
%
\begin{figure}
\epsfysize=4.2in
\centerline{\epsfbox{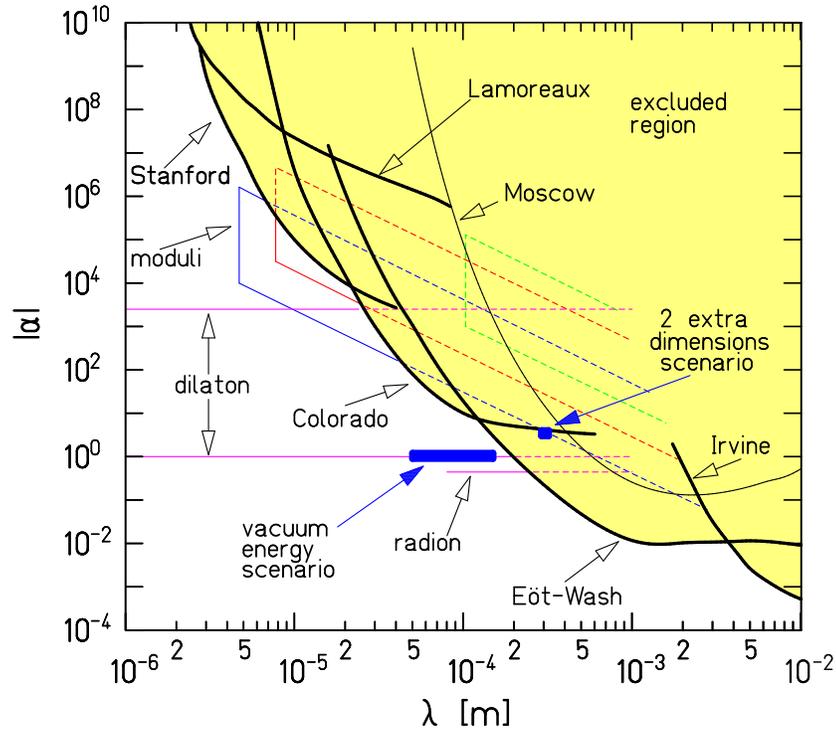}}
\caption{95\%-confidence-level constraints on ISL-violating Yukawa
interactions with 1~$\mu{\rm m} < \lambda < 1$ cm. The heavy curves
give experimental upper limits (the Lamoreaux constraint was computed
in Reference~\cite{lo:99}). Theoretical expectations for 
extra dimensions \cite{Ar:98a}, moduli \cite{DG:96}, dilaton \cite{Ka:00a}, 
and radion \cite{An:97} are shown as well.}
\label{fig:alpha lambda}
\end{figure}

\begin{figure}
\epsfysize=3.0in
\centerline{\epsfbox{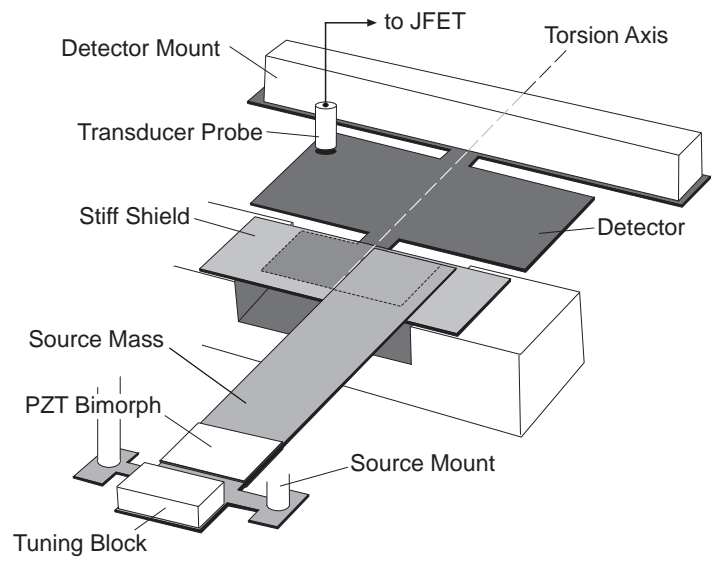}}
\caption{Schematic diagram of the instrument used by Long et al. \cite{lo:02}.}
\label{fig:long apparatus}
\end{figure}
%
\begin{figure}
\epsfysize=4.5in
\centerline{\epsfbox{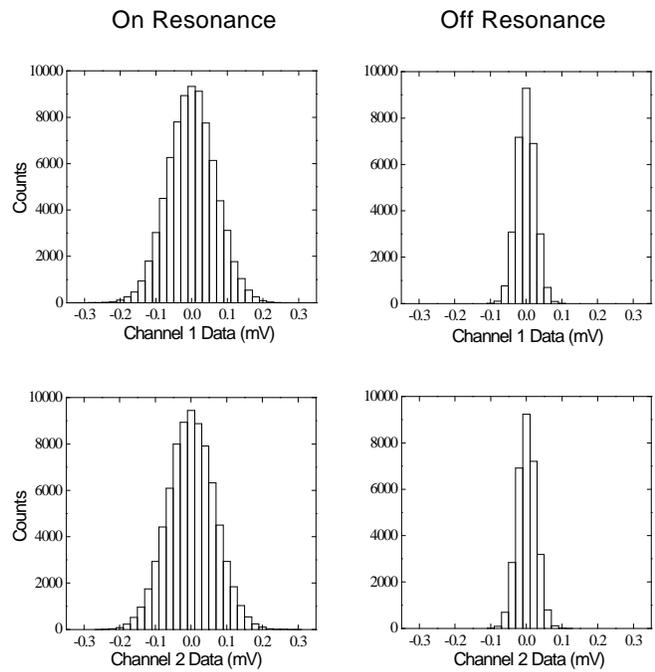}}
\caption{Data from the experiment of Long et al. \cite{lo:02} showing the two quadrature
signals from the torsion oscillator.}
\label{fig:long data}
\end{figure}

\begin{figure}
\epsfysize=2.2in
\centerline{\epsfbox{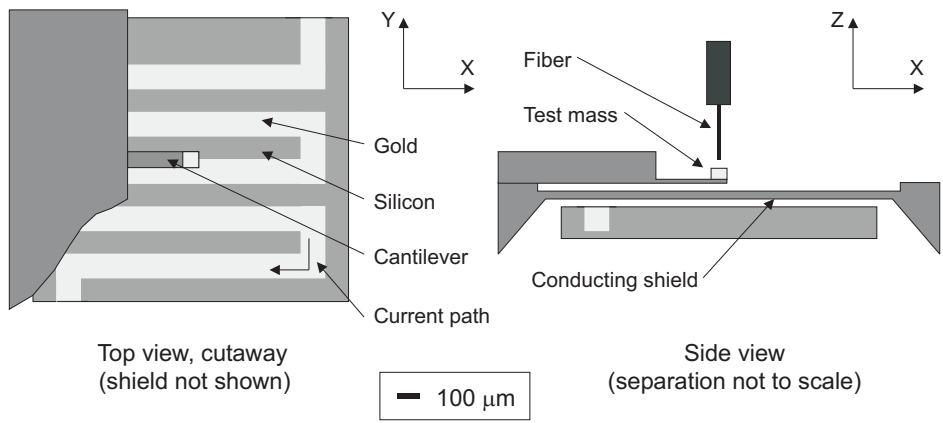}}
\caption{Schematic diagram of the instrument used by Chiaverini et al. \cite{ch:02}.}
\label{fig:kapitulnik instrument}
\end{figure}
%
\begin{figure}
\epsfysize=4.2in
\centerline{\epsfbox{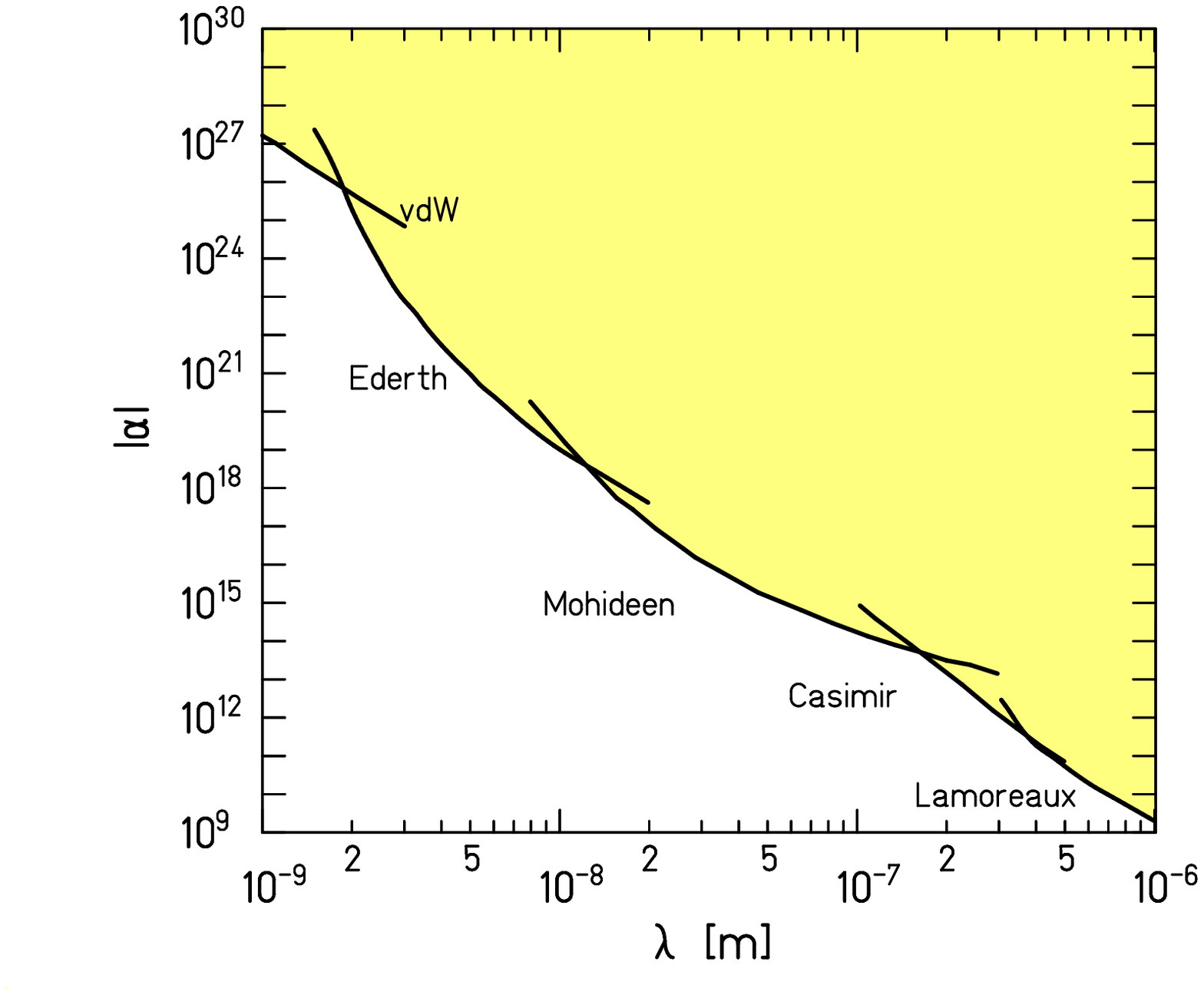}}
\caption{Constraints on ISL-violating Yukawa
interactions with $1{\rm nm}< \lambda < 1\mu$m adapted
from Reference~\cite{fi:01}. As discussed in the text, these upper limits,
extracted from Casimir-force measurements, are not as rigorous as
those in Figures~\ref{fig:alpha lambda} and \ref{fig:alpha lambda long}.}
\label{fig:alpha lambda short}
\end{figure}

\begin{figure}
\vspace* {0.2in}
\epsfysize=2.5in
\centerline{\epsfbox{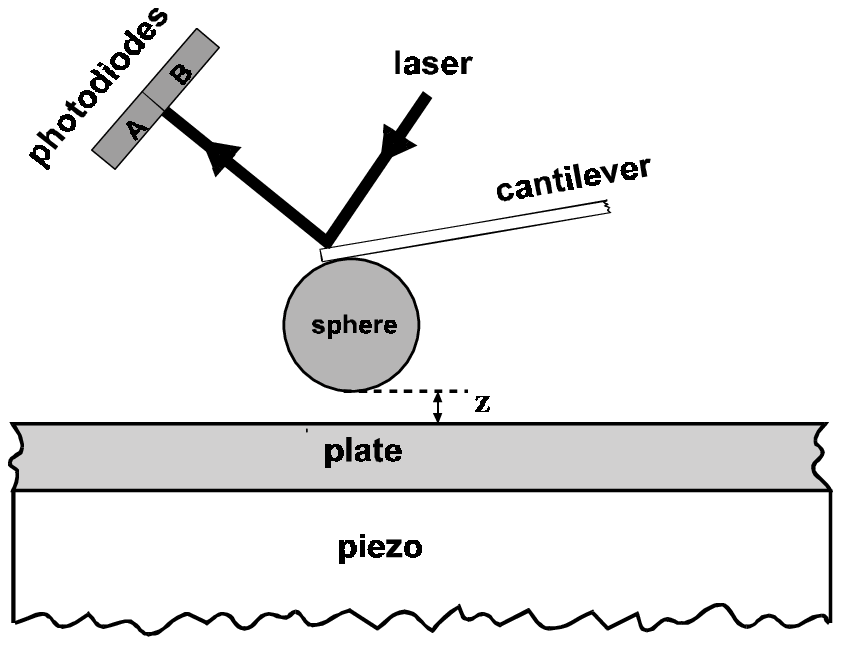}}
\caption{Schematic diagram of the Casimir-force apparatus used in
Reference~\cite{har:20}.}
\label{fig:afm}
\end{figure}

\begin{thebibliography}{99}
%
\bibitem{ri:98}Riess AG, et al., {\it Astron. J.} 116:1009 (1998)
%
\bibitem{pe:99}Perlmutter S, et al. {\it Astrophys. J.} 517:565 (1999)
%
\bibitem{coul}Erler J, Langacker P.
{\it Phys. Rev. D} 66:010001
(2002)
%
\bibitem{la:98}Lakes R. {\it Phys. Rev. Lett.} 80:1826 (1998)
%
\bibitem{wi:98}Will CM. {\it Phys. Rev. D} 57:2061 (1998)
%
\bibitem{newref_giu02}
Gruzinov A. astro-ph/0112246 (2002)
%
\bibitem{fi:01}Fischbach E, Krause DE, Mostepanenko VM, Novello M.
{\it Phys. Rev. D} 64:075010 (2001)
%
\bibitem{su:93}
Feinberg G, Sucher J. {\it Phys.\ Rev.\ D} 20:1717 (1979)
%
\bibitem{dr:53}Drell SD, Huang K. {\it Phys. Rev.} 91:1527 (1953)
%
\bibitem{mo:87a}Mostepanenko VM, Sokolov IYu. {\it Sov. J. Nucl. Phys.} 46:685
(1987)
%
\bibitem{fe:98}Ferrer F, Grifols JA. {\it Phys. Rev. D} 58:096006 (1998)
%
\bibitem{fi:96}Fischbach E. {\it Ann. Phys. (NY)} 247:213 (1996)
%
\bibitem{ad:91}Adelberger EG, Stubbs CW, Heckel BR, Rogers WF.
{\it Annu. Rev. Nucl. Part. Sci.} 41:269 (1991)
%
\bibitem{fi:99}Fischbach E, Talmadge CL. {\it The Search for Non-Newtonian
Gravity}. New York: Springer-Verlag (1999)
%
\bibitem{hew:02}Hewitt J, Spiropulu M.
{\it Annu. Rev. Nucl. Part. Sci.} 52:397 (2002)
%
%
\bibitem{newref_lon03}
Long JC, Price JC. hep-ph/0303057
%
\bibitem{Ge:76}
Gell-Mann M, Ramond P, Slansky R.
{\it Rev. Mod. Phys.} 50:721 (1978)
%
\bibitem{La:80}
Langacker P.
{\it Phys. Rep.} 72:185 (1981)
%
\bibitem{Sl:81}
Slansky R.
{\it Phys. Rep.} 79:1 (1981)
\bibitem{Ko:83}
Kolb EW, and Turner MS.
{\it Annu. Rev. Nucl. Part. Sci.} 33:645 (1983)
\bibitem{Ha:84}
Haber HE, Kane GL.
{\it Phys. Rep.} 117:75 (1985)
\bibitem{We:92}
Wess J, Bagger J.
{\it Supersymmetry and Supergravity}. Princeton, NJ: Princeton Univ.\ Press  (1992)
%
\bibitem{Dine:1996}
Dine M.
 {\it Boulder 1996, Fields, Strings and Duality.} Singapore: World Sci.\ (1997)
%
\bibitem{Da:97}
Dawson S. hep-ph/9712464,
 {\it Boulder 1997, Supersymmetry, Supergravity, and Super Colliders.} Singapore: World Sci.\ (1998)
%
\bibitem{Ba:96}
Bagger JA. 
In {\it Boulder 1995, QCD and Beyond.} Singapore: World Sci.\ (1996)
%
\bibitem{Lykken:1996}
Lykken JD.
In {\it Boulder 1996, Fields, Strings and Duality.} Singapore: World Sci.\ (1997)
%
\bibitem{Ma:97}
Martin SP.
In {\it Perspectives on Supersymmetry.} Singapore: World Sci.\ (1998)
%
\bibitem{Polonsky:2001}
Polonsky N.
{\it Lect.\ Notes Phys.}M68:1 (2001)
%
\bibitem{Pop:98}
Poppitz E, Trivedi SP.
{\it Annu.\ Rev.\ Nucl.\ Part.\ Sci.} 48:307 (1998)
%
\bibitem{Sh:99}
Shadmi Y, Shirman Y.
{\it Rev. Mod. Phys.} 72:25 (2000)
%
\bibitem{Ca:00}
Carroll SM.
{\it Living Rev. Rel.} 4:1 (2001)
%
\bibitem{Be:97}Beane SR.
{\it Gen.\ Rel.\ Grav.} 29:945 (1997)
arXiv:hep-ph/9702419.
%
\bibitem{Su:97}
Sundrum R.
{\it J. High Energy Phys.}   9907:001 (1999)
%
\bibitem{Be:93}
Bekenstein JD.
{\it Phys. Rev. D} 49:1912 (1994)
%
\bibitem{tH:93}
't Hooft G.
gr-qc/9310026
%
\bibitem{Su:94}
Susskind L.
{\it J. Math. Phys.} 36: 6377 (1995)
%
\bibitem{Bo:02}
Bousso R.
{\it Rev. Mod. Phys.} 74:825 (2002)
%
\bibitem{Ah:99}
Aharony O, et al.
{\it Phys. Rep.}  323:183 (2000)
%
\bibitem{Sc:82}
Schwarz JH.
{\it Phys. Rep.} 89:223 (1982)
%
\bibitem{Gr:87}
Green MB, Schwarz JH, Witten E.
{\it Superstring Theory.} Vol.\ 1, {\it Introduction};
Vol.\ 2, {\it Loop Amplitudes, Anomalies and Phenomenology.}
Cambridge, UK: Cambridge Univ.\ Press (1987)
%
\bibitem{Pol:98}
Polchinski J.
{\it String Theory.} Vol.\ 1, {\it An Introduction to the Bosonic String};
Vol.\ 2, {\it Superstring Theory and Beyond}. Cambridge, UK: Cambridge Univ.\ Press (1998)
%
\bibitem{Dienes:1996du}
Dienes KR.
{\it Phys. Rep.} 287:447 (1997)
%
\bibitem{Sc:96}
Schwarz JH.
{\it Nucl. Phys. Proc. Suppl.} 55B:1 (1997)
%
\bibitem{Du:96}
Duff MJ.
{\it Int. J. Mod. Phys. A} 11:5623 (1996)
%
\bibitem{Po:96a}
Polchinski J, Chaudhuri S, Johnson CV.
hep-th/9602052
%
\bibitem{Po:96b}
Polchinski J.
{\it Rev. Mod. Phys.} 68:1245 (1996)
%
\bibitem{Ki:97}
Kiritsis E.
In {\it Leuven Notes in Mathematical and Theoretical Physics}, B9. Leuven, Belgium: Leuven Univ.\ Press (1998)
%
\bibitem{Va:97}
Vafa C.
In {\it Trieste 1996, High Energy Physics and Cosmology.} Singapore: World Sci.\ (1997)
%
%
\bibitem{Gr:98}
Greene BR, Morrison DR, Polchinski J.
{\it Proc.\ Nat.\ Acad.\ Sci.} 95:11039 (1998)
%
\bibitem{Se:98}
Sen A.
In {\it Cambridge 1997, Duality and Supersymmetric Theories, Publications of Newton Institute.} Cambridge, UK: Cambridge Univ.\ Press (1999)
%
\bibitem{Ta:97}
Taylor WI.
In {\it Trieste 1997, High Energy Physics and Cosmology.} Singapore: World Sci.\ (1999)
%
\bibitem{Ba:98}
Bachas CP.
In {\it Cambridge 1997, Duality and Supersymmetric Theories.} Cambridge, UK: Cambridge Univ.\ Press (1999)
%
\bibitem{Di:99}
Di Vecchia P, Liccardo A.
hep-th/9912161, hep-th/9912275
%
\bibitem{Jo:00}
Johnson CV.
In {\it Boulder 1999, Strings, Branes, and Gravity.} Singapore: World Sci.\ (2000)
%
\bibitem{Dine:2000}
Dine M.
In {\it Boulder 1999, Strings, Branes, and Gravity.} Singapore: World Sci.\ (2000)
%
%
\bibitem{Ar:98a}
Arkani-Hamed N, Dimopoulos S, Dvali GR.
{\it Phys. Lett.} B429:263 (1998)
%
\bibitem{An:98}
Antoniadis I, Arkani-Hamed N, Dimopoulos S, Dvali GR.
{\it Phys. Lett.} B436:257 (1998)
%
\bibitem{Ly:96}
Lykken JD.
{\it Phys. Rev. D} 54:3693 (1996)
%
\bibitem{Ar:98b}
Arkani-Hamed N, Dimopoulos S, March-Russell J.
{\it Phys. Rev. D} 63:064020 (2001)
%
\bibitem{Ar:99a}
Arkani-Hamed N, Hall LJ, Smith DR, Weiner N.
{\it Phys. Rev. D} 62:105002 (2000)
%
\bibitem{Co:99}
Cohen AG, Kaplan DB.
{\it Phys. Lett.} B:470:52 (1999)
%
\bibitem{Ch:01}
Chacko Z, Fox PJ, Nelson AE, Weiner N.
{\it J. High Energy Phys.}  0203:001 (2002)
%
\bibitem{Al:01}
Albrecht A, Burgess CP, Ravndal F, Skordis C.
{\it Phys. Rev. D} 65:123506 (2002)
%
\bibitem{Ka:21}
Kaluza T.
{\it Sitzungsber.\ Preuss.\ Akad.\ Wiss.\ Berlin (Math.\ Phys.)} K1:966 (1921)
%
\bibitem{Kl:26a}
Klein O.
{\it Nature} 118:516 (1926)
%
\bibitem{Kl:26b}
Klein O.
{\it Z.\ Phys.} 37:895 (1926);
{\it Surveys High Energy. Phys.} 5:241 (1986)
%
\bibitem{Gi:98}
Giudice GF, Rattazzi R, Wells JD.
{\it Nucl. Phys. B} 544:3 (1999)
%
\bibitem{He:98}
Hewett JL.
{\it Phys. Rev. Lett.} 82:4765 (1999)
%
\bibitem{Ha:98}
Han T, Lykken JD, Zhang RJ.
{Phys. Rev. D} 59:105006 (1999)

\bibitem{Mi:98}
Mirabelli EA, Perelstein M, Peskin ME.
{\it Phys. Rev. Lett.} 82:2236 (1999)
%
\bibitem{ho:01}Hoyle CD, et al. {\it Phys. Rev. Lett.} 86:1418 (2001)
%
\bibitem{Ar:98c}
Arkani-Hamed N, Dimopoulos S, Dvali GR. {\it Phys. Rev. D}   59:086004 (1999)
%
\bibitem{Cu:99}
Cullen S, Perelstein M.
{\it Phys. Rev. Lett.} 83:268 (1999)
%
\bibitem{Ba:99}
Barger VD, Han T, Kao C, Zhang RJ.
{\it Phys. Lett.} B461:34 (1999)
%
\bibitem{Ha:00}
Hanhart C, Phillips DR, Reddy S, Savage MJ.
{\it Nucl.  Phys. B}  595:335 (2001)

\bibitem{Ha:01}
Hanhart C, Pons JA, Phillips DR. Reddy S.
{\it Phys. Lett.} B509:1 (2001)
%
\bibitem{Ha:99}
Hall LJ, Smith DR.
{\it Phys. Rev. D} 60:085008 (1999)
%
\bibitem{Ly:99b}
Lykken J, Nandi S.
 {\it Phys. Lett.} B485:224 (2000)
%
\bibitem{newref_bra61}
Brans C, Dicke QW. {\it {Phys.\ Rev.} }124:925 (1961)
%
\bibitem{Da:02}
Damour T. {\it Phys. Rev. D} 66:010001
(2002)
%
\bibitem{Ke:99}
Kehagias A, Sfetsos K.
{\it Phys.\ Lett.} B472:39 (2000)
%
\bibitem{Fl:99}
Floratos EG, Leontaris GK.
{\it Phys.\ Lett.} B465:95 (1999)
%
\bibitem{An:97}
Antoniadis I, Dimopoulos S, Dvali GR.
{\it Nucl. Phys. B} 516:70 (1998)
%
\bibitem{Ch:02}
Chacko Z, Perazzi E.
hep-ph/0210254
%
\bibitem{An:02}
Antoniadis I, Benakli K, Laugier A, Maillard T.
hep-ph/0211409. 
%
\bibitem{Ly:00}
Lykken JD.
In {\it TASI 2000, Flavor Physics for the Millennium.} Singapore: World Sci.\ (2001)
%
\bibitem{La:00}
Landsberg G.
hep-ex/0009038
%
\bibitem{Ru:01}
Rubakov VA.
{\it Phys.\ Usp.} 44:871 (2001);
{\it Usp.\ Fiz.\ Nauk }171:913 (2001)
%
\bibitem{He:02a}
Hewett J, March-Russell J.
{\it Phys. Rev. D} 66:010001
(2002)
%
\bibitem{Ue:02}
Uehara Y.
{\it Mod.\ Phys.\ Lett.\ A } 17:1551 (2002)
\bibitem{Ra:99a}
Randall L, Sundrum R.
{\it Phys. Rev. Lett.}  83:3370 (1999)
%
\bibitem{Go:99}
Goldberger WD, Wise MB.
{\it Phys. Rev. Lett.}   83:4922 (1999)
%
\bibitem{Fo:00}
Fox PJ.
{\it J. High Energy Phys.}  0102:022 (2001)
%
%
\bibitem{Ra:99b}Randall L, Sundrum R.
{\it Phys. Rev. Lett. } 83:4690 (1999)
%
\bibitem{Ar:99b}
Arkani-Hamed N, Dimopoulos S, Dvali  GR, Kaloper N.
{\it Phys. Rev. Lett.}  84:586 (2000)
%
\bibitem{Ly:99a}
Lykken J, Randall L.
{\it J. High Energy Phys.} 0006:014 (2000)
%
\bibitem{theta} Dine M.
In {\it TASI 2000, Flavor Physics for the Millennium.} Singapore: World Sci.\ (2001)
%
\bibitem{mo:84}Moody JE, Wilczek F. {\it Phys. Rev. D} 30:130 (1984)
%
%
\bibitem{Ve:99}
Verlinde E, Verlinde H.
{\it J. High Energy Phys.} 0005:034 (2000)
%
\bibitem{Sc:00}
Schmidhuber C.
{\it Nucl. Phys B} 580:140 (2000)
%
\bibitem{DG:96}
Dimopoulos S, Giudice GF.
{\it Phys. Lett.} B379:105 (1996)
%
\bibitem{Ka:00a}
Kaplan DB, Wise MB. {\it J. High Energy Phys.} 07:67 (2000)
%
%
\bibitem{Bars:86}
Bars I, Visser M.
{\it Phys.\ Rev.\ Lett.} 57:25 (1986)
%
\bibitem{It:86}
Itoyama H, McLerran LD, Taylor TR, van der Bij JJ.
{\it Nucl. Phys. B} 279:380 (1987)
%
\bibitem{Dv:00c}
Dvali GR, Gabadadze G, Porrati M.
{\it Mod.\ Phys.\ Lett.\ A }15:1717 (2000)
%
\bibitem{Dvali:2000hp}
Dvali GR, Gabadadze  G, Porrati M.
{\it Mod.\ Phys.\ Lett.\ A} 15:1717 (2000)
%
\bibitem{Barb:86}
Barbieri R, Cecotti S.
{\it Z.\ Phys.\ C} 33:255 (1986)
%
\bibitem{Barr:86}
Barr SM, Mohapatra RN.
{\it Phys. Rev. Lett.} 57:3129 (1986)
%
\bibitem{sm:00}
Smith GL, et al. {\it Phys. Rev. D} 61:022001
(2000)
%
%
\bibitem{We:88}
Weinberg S.
{\it Rev. Mod. Phys.} 61:1 (1989)
%
\bibitem{Ca:91}
Carroll SM, Press WH, Turner EL.
{\it Annu.\ Rev.\ Astron.\ Astrophys.} 30:499 (1992)
%
%
\bibitem{We:00}
Weinberg S.
In {\it  Int.\ Symp.\ Sources and Detection of Dark Matter in the Universe.} {Springer-Verlag (2001)
%
\bibitem{Wi:00}
Witten E.
 {\it Int.\ Symp.\ Sources and Detection of Dark Matter in the Universe.}  Springer-Verlag (2001)
%
\bibitem{Su:03}
Sundrum R. 
Presented at Conf.\ ``Frontiers beyond the Standard Model'' at Univ.\
Minn., Oct. 2002
%
\bibitem{Ru:83}
Rubakov VA, Shaposhnikov ME.
{\it Phys. Lett.} B125:139 (1983)
%
\bibitem{Gr:00}
Gregory R, Rubakov VA, Sibiryakov VM.
{\it Phys. Rev. Lett. } 84:5928 (2000)
%
\bibitem{Kar:00}
Karch A, Randall L.
{\it J. High Energy Phys.} 0105:008 (2001)
%
\bibitem{Dv:00a}
Dvali GR, Gabadadze G, Porrati M.
{\it Phys.\ Lett.}B485:208 (2000)
%
\bibitem{Dv:00b}
Dvali GR, Gabadadze G.
{\it Phys. Rev. D} 63:065007 (2001)
%
\bibitem{Ba:95}
Banks T.
hep-th/9601151
%
\bibitem{Ho:97}
Horava P.
{\it Phys. Rev. D} 59:046004 (1999)
%
\bibitem{Co:98}
Cohen AG, Kaplan DB, Nelson AE.
{\it Phys. Rev. Lett.} 82:4971 (1999)
%
\bibitem{Ba:00}
Banks T.
hep-th/0007146
%
\bibitem{Th:02}
Thomas S.
{\it Phys. Rev. Lett.} 89:081301 (2002)
%
\bibitem{Ba:02a}
Banks T.
hep-ph/0203066
%
\bibitem{Ba:02b}
Banks T.
hep-th/0206117
%
\bibitem{De:01}
Deffayet C, Dvali GR, Gabadadze G.
{\it Phys. Rev. D} 65:044023 (2002)
%
\bibitem{Dv:02a}
Dvali G, Gabadadze G, Shifman M.
hep-th/0202174
%
\bibitem{Dv:02b}
Dvali G, Gabadadze G, Shifman M.
hep-th/0208096
%
\bibitem{Ar:02}
Arkani-Hamed N, Dimopoulos S, Dvali G, Gabadadze G.
hep-th/0209227
%
\bibitem{Dvali:2001gm}
Dvali GR, Gabadadze G, Kolanovic M, Nitti F.
{\it Phys.\ Rev.\ D}  64:084004 (2001)
%
\bibitem{Deffayet:2001uk}
Deffayet C, Dvali GR, Gabadadze G, Vainshtein AI.
{\it Phys.\ Rev.\ D} 65:044026 (2002)
%
\bibitem{Gruzinov:2001hp}
Gruzinov A.
astro-ph/0112246
%
\bibitem{Lu:02}
Lue A, Starkman G.
astro-ph/0212083
%
\bibitem{Dv:02c}
Dvali G, Gruzinov A, Zaldarriaga M.
hep-ph/0212069
%
\bibitem{mi:88}
Mitrofanov VP, Ponomareva OI. {\it Sov. Phys. JETP} 67:1963
(1988) [translation of {\it Zh. Eksp. Teor. Fiz.} 94:16 (1988)]
%
\bibitem{ho:85}Hoskins JK, Newman RD, Spero R, Schultz J.
{\it Phys. Rev. D} 32:3084 (1985)
%
\bibitem{ed:00}Ederth T. {\it Phys. Rev. A} 62:062104 (2000)
%
\bibitem{la:97}Lamoreaux SK, {\it Phys. Rev. Lett.} 78:5 (1997)
%
\bibitem{umo:98}Mohideen U, Roy A. {\it Phys. Rev. Lett.} 81:4549 (1998)
%
\bibitem{lo:02}Long JC et al. hep-ph/0210004 (2002)
%
\bibitem{ha:00}Harris BW, Chen F, Mohideen U. {\it Phys. Rev. A}
62:052109 (2000)
%
\bibitem{sp:96}Speake CC. {\it Class. Quantum Grav.} 13:A291 (1996)
%
\bibitem{mi:77}Michaelson HB. {\it J. Appl. Phys.} 48:4729 (1977)
%
\bibitem{hbc:48}Casimir HBG. {\it Proc. K. Ned. Akad. Wet.} 51:793 (1948)
%
\bibitem{eml:56}Lifshitz EM. {\it Sov. Phys. JETP} 2:73 (1956)
%
\bibitem{gu:00}Gundlach JH, Merkowitz SM. {\it Phys. Rev. Lett.} 85:2869 (2000)
%
\bibitem{ho:01a}Hoyle CD. {\it Sub-millimeter Tests of the Gravitational
Inverse-Square Law}. PhD thesis. Univ. Wash. (2001)
%
\bibitem{he:02}
Adelberger EG. {\it Proc.\ Second Meeting on CPT and Lorentz Symmetry,} 
ed.\ VA Kostelecky, p.\ 9. Singapore: World Sci.\ (2002)
%
\bibitem{sa:90}Saulson PR. {\it Phys. Rev. D} 42:2437 (1990)
%
\bibitem{lo:99}Long JC, Chan HW, Price JC. {\it Nucl. Phys. B} 539:23 (1999)
%
\bibitem{ch:02}Chiaverini J et al. hep-ph/0209325 (2002)
%
\bibitem{ch:02a}Chiaverini J. {\it Small force detection using microcantilevers:
search for sub-millimeter-range deviations from Newtonian gravity.} PhD thesis. Stanford
Univ. (2002)
%
\bibitem{sp:58}Sparnaay MJ. {\it Physica (Amsterdam)} 24:751 (1958)
%
\bibitem{de:56}Derjaguin BV, Abrikosova II, Lifshitz EM. {\it Q. Rev. Chem. Soc.}
10:295 (1956)
%
\bibitem{ta:69}Tabor D, Winterton RHS. {\it Proc. R. Soc. London} A312:435 (1969)
%
\bibitem{is:72}Israelachvili YN, Tabor D. {\it Proc. R. Soc. London} A331:19 (1972)
%
\bibitem{hu:72}Hunklinger S, Geisselmann H, Arnold W. {\it Rev. Sci. Instrum.}
43:584 (1972)
%
\bibitem{ku:82}Kuz'min VA, Tkachev II, Shaposhnikov ME. {\it JETP Lett.} 36:59
(1982)
%
\bibitem{mo:87}Mostepanenko VM, Sokolov IYu. {\it Phys. Lett.} A125:405 (1987)
%
\bibitem{efi:92}Fischbach E, et al. {\it Metrologia} 29:215 (1992)
%
\bibitem{lam:98}Lamoreaux SK. {\it Phys. Rev. Lett} 81:5475 (1998)
%
\bibitem{roy:99}Roy A, Lin CY, Mohideen U. {\it Phys. Rev. D} 60:111101 (1999)
%
\bibitem{aro:99}Roy A, Mohideen U. {\it Phys. Rev. Lett.} 82:4380 (1999)
%
\bibitem{har:20}Harris BW, Chen F, Mohideen U. {\it Phys. Rev. A} 62:052109 (2000)
%
\bibitem{ede:20}Ederth T. {\it Phys. Rev. A} 62:062104 (2000)
%
\bibitem{wag:95}Wagner P, Hegner M, Guntherodt HJ, Semenza G. {\it Langmuir} 11:3867
(1995)
%
\bibitem{bor:98}Bordag M, Geyer B, Klimchitskaya GL, Mostepanenko VM. {\it Phys.
Rev. D} 58:075003 (1998)
%
\bibitem{kli:01}Klimchitskaya GL, Mostepanenko VM. {\it Phys. Rev. A} 63:062108
(2001) and references therein
%
\bibitem{gen:00}Genet C, Lambrecht A, and Reynaud S. {\it Phys. Rev. A} 62:012110 (2000)
%
\bibitem{bord:00}Bordag M, Geyer G, Klimchitskaya GL, Mostepanenko VM. {\it Phys.
Rev. Lett.} 85:503 (2000)
%
\bibitem{sch:78}Schwinger J, DeRaad LL, Milton KA. {\it Ann. Phys. (N.Y.)} 115:1
(1978)
%
\bibitem{bez:97}Bezerra VB, Klimchitskaya GL, Romero C. {\it Mod. Phys. Lett. A}
12:2613 (1997)
%
\bibitem{klim:99}Klimchitskaya GL, Roy A, Mohideen U, Mostepanenko VM. {\it Phys.
Rev. A} 60:3487 (1999)
%
\bibitem{lamo:99}Lamoreaux SK, {\it Phys. Rev A} 59:R3149 (1999)
%
\bibitem{lamb:00}Lambrecht A, Reynaud S. {\it Eur. Phys. J. D} 8:309 (2000)
%
\bibitem{bos:00}Bostrom M, Sernelius BoE. {\it Phys. Rev. A} 61:046101 (2000)
%
\bibitem{vb:74}Van Bree JLMM, Poulis JA, Verhaar BJ, Schram K. 
{\it Physics (Amsterdam)} 78:187 (1974)
%
\bibitem{mara:80}Maradudin AA, Mazur P. {\it Phys. Rev. B} 22:1677 (1980); 23:695
(1981)
%
\bibitem{klim:96}Klimchitskaya GL, Pavlov YuV. {\it Int. J. Mod. Phys. A} 11:3723
(1996)
%
\bibitem{most:01}Mostepanenko VM, Novello M. {\it Phys. Rev. D} 63:115003 (2001)
%
\bibitem{bord:99}Bordag M, Geyer G, Klimchitskaya GL, Mostepanenko VM. {\it Phys.
Rev. D} 60:055004 (1999)
%
\bibitem{bor:00}Bordag M, Geyer G, Klimchitskaya GL, Mostepanenko VM.{\it Phys.
Rev. D} 62:011701R (2000)
%
\bibitem{mos:00}Mostepanenko VM, Novello M. hep-ph/0008035 v1 (2000)
%
\bibitem{bor:94}Bordag M, Mostepanenko VM, Sokolov IYu. {\it Phys. Lett.} A187:35
(1994)
%
\bibitem{llrgeneral}Williams JG, Newhall XX, Dickey JO. {\it Phys. Rev. D} 53:6730
(1996)
%
\bibitem{sh:85}Shelus PJ. {\it IEEE Trans. Geosci. Remote Sensing} GE-23:385 (1985)
%
\bibitem{sa:98}Samain E et al. {\it Astron. Astrophys. Suppl. Ser.} 130:235 (1998)
%
\bibitem{wi:02}Williams JG, Boggs DH, Dickey JO, Folkner WM.
{\it Proc. 9th Marcel Grossmann Meeting} Rome 2002, p.\ 1797. Singapore: World  Sci.
%
\bibitem{ba:00}Bantel MK, Newman RD. {\it J. Alloys Compounds} 310:233 (2000)
%
\bibitem{mu:00}Murphy T et al.
{\it The Apache Point Observatory lunar laser-ranging operation. }
{  http://geodaf.mt.asi.it/GDHTL/news/iwlr/Murphy{\_}et{\_}al{\_}apachepoint.pdf } 
(2000) 
%
\bibitem{de:02}Degnan JJ. {\it J. Geodynamics} 34:551 (2002)
%
\bibitem{no:}
Anderson JD, Gross M, Nordtvedt K, Turyshev S. {\it Astrophys. J.} 459:365
(1996)
}\end{thebibliography}
\end{document}